\documentclass{aastex63}

\received{}
\revised{}
\accepted{\today}
\submitjournal{AAS Journals}

\shorttitle{Pluto--Charon Sonata IV}
\shortauthors{Kenyon \& Bromley}

\usepackage{epsf,amssymb,amsmath}
\usepackage{wrapfig}

\def\nbody{$n$-body}

\def\deg{\ifmmode {^\circ}\else {$^\circ$}\fi}
\def\degree{\ifmmode {^\circ}\else {$^\circ$}\fi}
\def\mum{\ifmmode {\rm \,\mu {\rm m}}\else $\rm \,\mu {\rm m}$\fi}
\def\arcsec{\ifmmode ^{\prime \prime}\else $^{\prime \prime}$\fi}

\def\inch{\ifmmode ^{\prime \prime}\else $^{\prime \prime}$\fi}
\def\gs{\ifmmode {{\rm g~s^{-1}}}\else ${\rm g~s^{-1}}$\fi}
\def\msunyr{\ifmmode {M_{\odot}~{\rm yr^{-1}}}\else $M_{\odot}~{\rm yr^{-1}}$\fi}
\def\msun{\ifmmode {M_{\odot}}\else $M_{\odot}$\fi}
\def\rsun{\ifmmode {R_{\odot}}\else $R_{\odot}$\fi}
\def\lsun{\ifmmode {L_{\odot}}\else $L_{\odot}$\fi}
\def\mstar{\ifmmode {M_{\star}}\else $M_{\star}$\fi}
\def\rstar{\ifmmode {R_{\star}}\else $R_{\star}$\fi}
\def\tstar{\ifmmode {T_{\star}}\else $T_{\star}$\fi}
\def\lstar{\ifmmode {L_{\star}}\else $L_{\star}$\fi}
\def\mwd{\ifmmode {M_{wd}}\else $M_{wd}$\fi}
\def\rwd{\ifmmode {R_{wd}}\else $R_{wd}$\fi}
\def\twd{\ifmmode {T_{wd}}\else $T_{wd}$\fi}
\def\lwd{\ifmmode {L_{wd}}\else $L_{wd}$\fi}
\def\md{\ifmmode {M_d}\else $M_d$\fi}
\def\ld{\ifmmode {L_d}\else $L_d$\fi}
\def\ad{\ifmmode A_d\else $A_d$\fi}
\def\ldlwd{\ifmmode L_d / L_{wd}\else $L_d / L_{wd}$\fi}
\def\ldlstar{\ifmmode L_d / L_\star\else $L_d / L_{\star}$\fi}
\def\rearth{\ifmmode {\rm R_{\oplus}}\else $\rm R_{\oplus}$\fi}
\def\mearth{\ifmmode {\rm M_{\oplus}}\else $\rm M_{\oplus}$\fi}
\def\qc{\ifmmode Q_c\else $Q_c$\fi}
\def\qdstar{\ifmmode Q_D^\star\else $Q_D^\star$\fi}
\def\rt{\ifmmode r_t\else $r_t$\fi}
\def\vc{\ifmmode v_c\else $v_c$\fi}
\def\vsqd{\ifmmode v^2 / Q_D^\star\else $v^2 / Q_D^\star$\fi}
\def\kms{\ifmmode {\rm km~s^{-1}}\else $\rm km~s^{-1}$\fi}
\def\ms{\ifmmode {\rm m~s^{-1}}\else $\rm m~s^{-1}$\fi}
\def\vrel{\ifmmode v_{rel}\else $v_{rel}$\fi}
\def\mdot{\ifmmode \dot{M}\else $\dot{M}$\fi}
\def\mdotz{\ifmmode \dot{M}_0\else $\dot{M}_0$\fi}
\def\mesc{\ifmmode m_{esc}\else $m_{esc}$\fi}
\def\rmin{\ifmmode r_{min}\else $r_{min}$\fi}
\def\rmax{\ifmmode r_{max}\else $r_{max}$\fi}
\def\xmax{\ifmmode x_{max}\else $x_{max}$\fi}
\def\mmin{\ifmmode m_{min}\else $m_{min}$\fi}
\def\mmax{\ifmmode m_{max}\else $m_{max}$\fi}
\def\rmind{\ifmmode r_{min,d}\else $r_{min,d}$\fi}
\def\rmaxd{\ifmmode r_{max,d}\else $r_{max,d}$\fi}
\def\mmaxd{\ifmmode m_{max,d}\else $m_{max,d}$\fi}
\def\vrad{\ifmmode v_{rad}\else $v_{rad}$\fi}
\def\qz{\ifmmode q_{0}\else $q_{0}$\fi}
\def\qi{\ifmmode q_{i}\else $q_{i}$\fi}
\def\ql{\ifmmode q_{l}\else $q_{l}$\fi}
\def\qs{\ifmmode q_{s}\else $q_{s}$\fi}
\def\vhill{\ifmmode v_H\else $r_H$\fi}
\def\egc{\ifmmode r_{gc}\else $r_{gc}$\fi}
\def\rgc{\ifmmode r_{gc}\else $r_{gc}$\fi}
\def\qgc{\ifmmode q_{gc}\else $q_{gc}$\fi}
\def\Qgc{\ifmmode Q_{gc}\else $Q_{gc}$\fi}
\def\rhill{\ifmmode r_H\else $r_H$\fi}
\def\Rhill{\ifmmode R_H\else $R_H$\fi}
\def\rbrk{\ifmmode r_{brk}\else $r_{brk}$\fi}
\def\rdamp{\ifmmode r_{damp}\else $r_{damp}$\fi}
\def\rin{\ifmmode r_{in}\else $r_{in}$\fi}
\def\rout{\ifmmode r_{out}\else $r_{out}$\fi}
\def\tin{\ifmmode t_{in}\else $t_{in}$\fi}
\def\tout{\ifmmode t_{out}\else $t_{out}$\fi}
\def\ain{\ifmmode a_{in}\else $a_{in}$\fi}
\def\aout{\ifmmode a_{out}\else $a_{out}$\fi}
\def\ag{\ifmmode a_{g}\else $a_{g}$\fi}
\def\eg{\ifmmode e_{g}\else $e_{g}$\fi}
\def\qg{\ifmmode q_{g}\else $q_{g}$\fi}
\def\Qg{\ifmmode Q_{g}\else $Q_{g}$\fi}
\def\ageo{\ifmmode a_{geo}\else $a_{geo}$\fi}
\def\egeo{\ifmmode e_{geo}\else $e_{geo}$\fi}
\def\qgeo{\ifmmode q_{geo}\else $q_{geo}$\fi}
\def\Qgeo{\ifmmode Q_{geo}\else $Q_{geo}$\fi}
\def\abin{\ifmmode a_{bin}\else $a_{bin}$\fi}
\def\ebin{\ifmmode e_{bin}\else $e_{bin}$\fi}
\def\efree{\ifmmode e_{free}\else $e_{free}$\fi}
\def\r0{\ifmmode r_{0}\else $r_{0}$\fi}
\def\R0{\ifmmode R_{0}\else $R_{0}$\fi}
\def\m0{\ifmmode m_{0}\else $m_{0}$\fi}
\def\mone{\ifmmode m_{1}\else $m_{1}$\fi}
\def\mtwo{\ifmmode m_{2}\else $m_{2}$\fi}
\def\atwo{\ifmmode a_{2}\else $a_{2}$\fi}
\def\etwo{\ifmmode e_{2}\else $e_{2}$\fi}
\def\mf{\ifmmode m_{f}\else $m_{f}$\fi}
\def\af{\ifmmode a_{f}\else $a_{f}$\fi}
\def\ef{\ifmmode e_{f}\else $e_{f}$\fi}
\def\M0{\ifmmode M_{0}\else $M_{0}$\fi}
\def\amax{\ifmmode a_{max}\else $a_{max}$\fi}
\def\a0{\ifmmode a_{0}\else $a_{0}$\fi}
\def\e0{\ifmmode e_{0}\else $e_{0}$\fi}
\def\v0{\ifmmode v_{0}\else $v_{0}$\fi}
\def\xm{\ifmmode x_{m}\else $x_{m}$\fi}
\def\sigz{\ifmmode \Sigma_0\else $\Sigma_0$\fi}
\def\ergg{\ifmmode {\rm erg~g^{-1}}\else ${\rm erg~g^{-1}}$\fi}
\def\ergs{\ifmmode {\rm erg~s^{-1}}\else ${\rm erg~s^{-1}}$\fi}
\def\gyr{\ifmmode {\rm g~yr^{-1}}\else ${\rm g~yr^{-1}}$\fi}
\def\cms{\ifmmode {\rm cm~s^{-1}}\else ${\rm cm~s^{-1}}$\fi}
\def\gcms{\ifmmode {\rm g~cm^{-2}}\else $\rm g~cm^{-2}$\fi}
\def\gcmc{\ifmmode {\rm g~cm^{-3}}\else $\rm g~cm^{-3}$\fi}
\def\atil{\ifmmode {\tilde{a}}\else $\tilde{a}$\fi}
\def\ttil{\ifmmode {\tilde{t}}\else $\tilde{t}$\fi}
\def\sqrttt{\ifmmode {\tilde{t}^{1/2}}\else $\tilde{t}^{1/2}$\fi}

\def\orch{{\it Orchestra}}
\def\nh{{\it New Horizons}}
\def\pc{Pluto--Charon}
\def\mp{\ifmmode m_P\else $m_P$\fi}
\def\mc{\ifmmode m_C\else $m_C$\fi}
\def\mh{\ifmmode m_H\else $m_H$\fi}
\def\mk{\ifmmode m_K\else $m_K$\fi}
\def\ms{\ifmmode m_S\else $m_S$\fi}
\def\mn{\ifmmode m_N\else $m_N$\fi}
\def\Rp{\ifmmode R_P\else $R_P$\fi}
\def\rp{\ifmmode r_P\else $r_P$\fi}
\def\rc{\ifmmode r_C\else $r_C$\fi}
\def\apc{\ifmmode a_{PC}\else $a_{PC}$\fi}
\def\mpc{\ifmmode m_{PC}\else $m_{PC}$\fi}
\def\epc{\ifmmode e_{PC}\else $e_{PC}$\fi}

\begin{document}

\title{A Pluto--Charon Sonata IV. Improved Constraints on the Dynamical Behavior and Masses of the Small Satellites}

\correspondingauthor{Scott J. Kenyon}
\email{skenyon@cfa.harvard.edu}

\author[0000-0003-0214-609X]{Scott J. Kenyon}
\affil{Smithsonian Astrophysical Observatory,
60 Garden Street,
Cambridge, MA 02138, USA}

\author[0000-0001-7558-343X]{Benjamin C. Bromley}
\affil{Department of Physics \& Astronomy,
University of Utah, 201 JFB,
Salt Lake City, DC 20006, USA}

\begin{abstract}
We discuss a new set of $\sim$ 500 numerical \nbody\ calculations 
designed to constrain the masses and bulk densities of Styx, Nix, 
Kerberos, and Hydra.  Comparisons of different techniques for deriving 
the semimajor axis and eccentricity of the four satellites favor methods 
relying on the 
theory of \citet{lee2006}, where satellite orbits are derived in
the context of the restricted three body problem (Pluto, Charon, and
one massless satellite).
In each simulation, we adopt the nominal satellite masses derived 
in \citet{kb2019b}, multiply the mass of at least one satellite by a 
numerical factor $f \ge 1$, and establish whether the system ejects at 
least one satellite on a time scale $\le$ 4.5~Gyr. 
When the total system mass is large ($f \gg 1$), ejections of Kerberos 
are more common. Systems with lower satellite masses ($ f \approx$ 1)
usually eject Styx.
In these calculations, Styx often
`signals' an ejection by moving to higher orbital inclination long
before ejection; Kerberos rarely signals in a useful way.
The \nbody\ results suggest that Styx and Kerberos are more likely 
to have bulk densities comparable with water ice, 
$\rho_{SK} \lesssim$ 2~\gcmc, than with rock.  A strong upper limit on 
the total system mass, $M_{SNKH} \lesssim 9.5 \times 10^{19}$~g, also 
places robust constraints on the average bulk density of the four 
satellites, $\rho_{SNKH} \lesssim$ 1.4~\gcmc. These limits support
models where the satellites grow out of icy material ejected
during a major impact on Pluto or Charon.
\end{abstract}



\section{Introduction} \label{sec: intro}

In the past two decades, space observations added new insights into the 
properties of the dwarf planet Pluto. From 2005--2012, HST images revealed 
four small circumbinary satellites \citep{weaver2006,showalter2011,
showalter2012}. Detailed astrometric analyses of these data demonstrate 
that the orbits of the central \pc\ binary and the satellites are nearly 
circular and in a common plane \citep{buie2006,tholen2008,brozovic2015,
showalter2015}.  Spectacular observations acquired during the \nh\ flyby 
confirm that the satellites tumble with approximate rotation periods of 
0.43~d to 5.31~d \citep{showalter2015,weaver2016}.  All of the satellites 
are irregularly shaped and highly reflective. Characteristic radii are 
$\sim$ 5~km for Styx and Kerberos and $\sim$ 20~km for Nix and Hydra. 
Albedos are $\sim$ 55\% for Kerberos and Nix, 65\% for Styx, and 85\% 
for Hydra \citep{weaver2016}. Although smaller satellites could exist 
slightly inside the orbit of Styx and outside the orbit of Hydra 
\citep{kb2019a}, there are no $\gtrsim$ 2~km satellites and a negligible 
amount of dust between the orbits of Styx and Hydra \citep{weaver2016, 
lauer2018}.

Deriving limits on the masses and the bulk densities of the small 
satellites requires detailed \nbody\ calculations \citep[e.g.,][and 
references therein]{pires2011,youdin2012,canup2021}. 
To improve on mass limits inferred from HST observations 
\citep{brozovic2015},
\citet{kb2019b} performed a large suite of \nbody\ calculations for various
combinations of satellite masses. Within the set of completed simulations,
they show that a `heavy' satellite system -- where the mass of Kerberos 
is roughly one third the mass of Hydra and the total system mass is 
$M_{SNKH} \sim 1.15 \times 10^{20}$~g \citep{brozovic2015} -- is unstable 
%
on time scales $\lesssim$ 1~Gyr.  For a `light' satellite system -- where
Nix/Hydra have the masses derived by \citet{brozovic2015} and Styx/Kerberos
have masses $\sim$ 10--25 times smaller -- the analysis yields
firm upper limits on the masses of Nix and Hydra.  Combined 
with physical dimensions measured from \nh\ images, the resulting upper 
limits on the bulk densities are $\rho_N \lesssim$ 1.3--1.6~\gcmc\ for 
Nix and $\rho_H \lesssim$ 1.1--1.5~\gcmc\ for Hydra. Both upper limits 
lie below the measured bulk densities for Pluto $\rho_P$ = 1.85~\gcmc\ 
and Charon $\rho_C$ = 1.70~\gcmc. 

Although the completed simulations in \citet{kb2019b} considerably reduced 
upper limits on the masses of Styx and Kerberos, upper limits on the masses
of Nix and Hydra relied on a mixture of completed and unfinished calculations. 
Using long-term trends in the evolution of a basic `geometric' eccentricity 
\citep[section 2; see also][]{sutherland2019} in many unfinished calculations,
\citet{kb2019b} placed stronger limits on the masses of Nix and Hydra. While 
these trends were robust among all the unfinished calculations, periods of
a steadily increasing eccentricity might be a temporary feature of a system 
that fails to eject satellites over the 4.5~Gyr lifetime of the solar system.

Here, we describe a new analysis of a larger set of completed 
simulations, supplemented with insights gleaned from several ongoing 
calculations. Adopting nominal (low) masses for Styx and Kerberos, 
calculations where at least one satellite has been ejected confirm 
previous upper limits on the masses of Nix and Hydra. Including the small 
nominal masses of Styx and Kerberos, a more reliable upper limit
on the system mass is $M_{SNKH} \lesssim 9.5 \times 10^{19}$~g.  
Coupled with satellite volumes estimated from \nh\ measurements of 
satellite dimensions, the average bulk density of a satellite 
$\rho_{SNKH} \lesssim$ 1.4~\gcmc.  

Another set of completed calculations begins to place limits on the
masses of Styx and Kerberos. Analysis of simulations where Styx and 
Kerberos have bulk densities of 2--3~\gcmc\ suggest significantly shorter 
lifetimes than a parallel suite of calculations where the bulk densities
are 1.0--1.5~\gcmc. Given the robust upper limits on the masses for Nix 
and Hydra, these results enable 
approximate upper limits for their bulk densities. While not as stringent
as the conclusions for Nix and Hydra, the calculations favor lower bulk 
densities, $\rho_{SK} \lesssim$ 2~\gcmc. This result suggests that these 
two small satellites are more likely composed of ice than rock, as 
indicated by their high albedos \citep{weaver2016}.

Besides improvements in limits on satellite masses, the \nbody\ 
calculations reveal interesting dynamical behavior. 
In systems with more (less) massive satellites, 
Kerberos (Styx) is ejected more often than Styx (Kerberos). 
Ejections of both are very rare.  Although Nix is never ejected, 
Hydra is sometimes ejected when the system mass is large. 
Among the systems that experience an ejection of Hydra, the frequency of 
Styx ejections is similar to the frequency of Kerberos ejections. 
During the period before an ejection, Styx often exhibits an oscillation 
where growth in its eccentricity is followed by a rise in inclination; 
increases in inclination are accompanied by a decline in eccentricity.
A later rise in eccentricity leads to a prompt ejection. Kerberos almost 
never follows this type of evolution.

In addition to the analyses described here and in \citet{kb2019b}, we deposit
binary files from all completed \nbody\ calculations and some of the programs 
used to extract and analyze the phase space coordinates at a publicly accessible 
repository (https://hive.utah.edu/). The combined set of 700 files from 
\citet{kb2019b} and 500 files from this study provides interested 
researchers a large data set for other analyses. 

In the next section, we outline the initial conditions for each calculation
and the numerical procedure. We then describe the results and discuss their
significance. We conclude with a brief summary.

\section{Calculations} \label{sec: calcs}

\subsection{Procedures}

We perform numerical calculations with a gravitational $n$-body code which
integrates the orbits of Pluto, Charon, and the four smaller satellites in
response to their mutual gravitational interactions \citep[e.g.,][]{kb2019a,
kb2019b,kb2019c}.  The $N$-body code, \orch, employs an adaptive sixth-order 
accurate algorithm based on either Richardson extrapolation \citep{bk2006} or 
a symplectic method \citep{yoshida1990,wisdom1991,saha1992}.  The code 
calculates gravitational forces by direct summation and evolves particles 
accordingly in the center-of-mass frame.  The code has passed a stringent set 
of dynamical tests and benchmarks \citep{dunc1998,bk2006}. 
\citet{bk2020} and \citet{kb2021a} describe recent improvements to the code
and cite additional tests of the algorithm.

The calculations do not include tidal or radiation pressure forces on the 
satellites \citep[e.g.,][]{burns1979,hamilton1992,poppe2011,pires2013,
quill2017}. Radiation pressure forces are significant on dust grains, but
satellites with sizes similar to and larger than Styx and Kerberos are 
unaffected. With a fixed orbit for the central binary, tidal forces have 
little impact on satellite orbits.

During the symplectic integrations, there is no attempt to resolve 
collisions between the small satellites or between an ejected satellite 
and Pluto or Charon. Satellites passing too close to another massive 
object in the system are eventually ejected. In the adaptive integrator,
the code changes the length of timesteps to resolve collisions.  In 
agreement with previous results \citep{sutherland2016,smullen2016,
smullen2017}, small satellites are always ejected from the system and 
never collide with other small satellites, Charon, or Pluto.

Previous studies demonstrate that the orbits of the small satellites 
are too far inside the Hill sphere of Pluto to require including the 
gravity of the Sun or major planets in the integrations 
\citep{michaely2017}. For reference, the radius of the Pluto-Charon Hill 
sphere is $R_{H,PC} \approx 8 \times 10^6$~km
for masses \mp\ = $1.303 \times 10^{25}$~g (Pluto) and 
\mc\ = $1.587 \times 10^{24}$~g (Charon).
In Hill units, the semimajor axis of Hydra's orbit,
$a_H / R_{H,PC} \approx$ 0.008, 
is well inside the Hill sphere and fairly immune 
from the gravity of the Sun.  
For the calculations described in this paper, the \nbody\ code 
follows the orbits of \pc\ and the four small satellites without any 
contribution from the gravity of the Sun or major planets. Previous 
tests with the \orch\ code show that including the Sun and the major 
planets has no impact on satellite orbits \citep{kb2019b}.

Throughout the \nbody\ calculations, we record the 6D cartesian phase space
variables, the Keplerian semimajor axis $a_K$ and eccentricity $e_K$, and the 
orbital inclination $\imath$ at the end of selected time steps.  Over total 
integration times as long as 0.1--2~Gyr, a typical calculation has 30,000 to 
more than 100,000 of these `snapshots' of the satellite positions, velocities, 
and Kplerian orbital parameters at machine precision.  To avoid unwieldy data 
sets, we make no attempt to record satellite positions during each orbit. 
Within the circumbinary environment of \pc, satellite orbits precess on time 
scales ranging from 1.2~yr for Styx to 2.8~yr for Hydra 
\citep[e.g.,][]{lee2006,leung2013,bk2015a}.  For any calculation, the ensemble 
of snapshots is insufficient to track the precession of the small satellites.

On the NASA `discover' cluster, 24~hr integrations on a single processor advance
the satellite system $\sim$ 4.3~Myr.  We perform 28 calculations per node, with
each satellite system evolving on one of the 28 cores per node.  To derive results
for as many sets of initial conditions as possible, the suite of simulations uses
6--10 nodes each day. In this way, each system advances $\sim$ 125~Myr per month.

\begin{deluxetable}{lccccccc}
\tablecolumns{8}
\tabletypesize{\footnotesize}
\tablenum{1}
\tablecaption{Adopted Masses and Initial Conditions}
\tablehead{
\colhead{Satellite} &
\colhead{Mass (g)} &
  \colhead{$x$ (km)} &
  \colhead{$y$ (km)} &
  \colhead{$z$ (km)} &
  \colhead{$v_x$ (km~s$^{-1}$)} &
  \colhead{$v_y$ (km~s$^{-1}$)} &
  \colhead{$v_z$ (km~s$^{-1}$)}
}
\label{tab: init}
\startdata
Pluto & $1.303 \times 10^{25}$ & -157.8121679944 & -456.7988459683 & -2071.4067337364 & -0.0177032091 & -0.0158015359 & 0.0048362971 \\
Charon & $1.586 \times 10^{24}$ & 1297.1743847853 &  3752.6022617472 & 17011.9058384535 & 0.1453959509 & 0.1297771902 & -0.0397230040 \\
Styx & $6 \times 10^{17}$ & -30572.8427772584 & -26535.8134344897 & 12311.2908958766 & 0.0232883189 & 0.0427977975 & 0.1464990284 \\
Nix & $4.5 \times 10^{19}$ & 9024.3487802378 & 15210.7370165008 & 45591.7573572213 & 0.1004334400  &  0.0865524814 & -0.0479498746 \\
Kerberos & $9 \times 10^{17}$ & 23564.2070250521 & 28380.0399507624 & 44578.0258218278 & 0.0792537026 & 0.0630220100 & -0.0817084451 \\
Hydra & $4.8 \times 10^{19}$ & -43331.3261132443 & -43628.4575945387 & -20506.5419357332 & -0.0374001038 & -0.0184905611 & 0.1157937283 \\
\enddata
\end{deluxetable}

\subsection{Initial Conditions}

All calculations begin with the same {\it measured} initial state vector
\citep{brozovic2015} for the 3D cartesian position -- $\vec{r} = (x, y, z)$ --
and velocity -- $ \vec{v} = (v_x, v_y, v_z)$ -- of each component.
Tests with a state vector downloaded from the JPL Horizons
website\footnote{https://ssd.jpl.nasa.gov/horizons.cgi} yield indistinguishable
results. \citet{kb2019b} describe several procedures for deriving the initial
state vector for Pluto, which is not included in \citet{brozovic2015}. For the
satellite masses considered here, we employ the Pluto-2 state vector and the
state vectors for the satellites listed in Table~\ref{tab: init} 
\citep[][see also, their Table 2]{kb2019b}. Test
calculations show that outcomes are insensitive to modest changes -- 0.5~km in 
position and 1.0~cm~s$^{-1}$ in velocity -- to the state vectors.

Although all calculations begin with the same initial state vector 
for Pluto, Charon, and the four small satellites, we perform each 
simulation with different satellite masses. We first adopt the nominal 
masses for Styx, \ms\ = 0.6; Nix, \mn\ = 45; Kerberos, \mk\ = 0.9; and 
Hydra, \mh\ = 48 in units of $10^{18}$~g (Table~\ref{tab: init}).
In some calculations, we multiply the nominal masses for each satellite 
by a factor $f = n (1 + \delta)$, where $n$ is an integer or simple 
fraction (e.g., 0.5, 0.625, 0.75, 0.875, 1.25 or 1.5) and $\delta$ is a 
small real number in the range $-$0.01 to 0.01. 
For a suite of calculations with similar $f$, $n$ and $\delta$ are 
{\it the same for all satellites}.  In other simulations, we multiply the 
mass of a single satellite by a factor $f_i$ and set the masses of the 
remaining satellites at their nominal masses. 

To avoid confusion, we use $f$ as a marker for calculations where we 
multiply the masses of all satellites by a common factor and $f_i$ (where 
$i$ = `S' for Styx,
`N' for Nix, `K' for Kerberos, and `H' for Hydra) as markers where 1--2 satellites
have masses that differ from the nominal masses. In some calculations, we set
$f_S = f_K$ = 1.5, 2, or 3 and then multiply masses for all four satellites by
a common $f$. In these models, Styx and Kerberos have masses $f_S \times f$ 
larger than their nominal masses.

For systems where all satellite masses have the same $f$, the $\delta$ term allows 
measurement of a range of lifetimes for systems with identical initial positions 
and velocities and nearly identical masses.  In many marginally stable dynamical 
systems, lifetimes are highly sensitive to initial conditions. Rather than make 
slight modifications to the adopted state vector to test this sensitivity, we use 
the $\delta$ term in the expression for $f$. As we showed in \citet{kb2019b}, 
1\% variations in $f$ result in factor of 3--10 differences in derived lifetimes.

Before starting the suite of calculations reported here, we considered adopting a 
volume for each satellite based on \nh\ measurements and deriving system stability 
as a function of satellite bulk density instead of mass. Although the 
\nh\ size 
measurements are nominally more accurate than the HST mass estimates, satellite 
shapes are not precisely known.  For an adopted shape, the $\pm$2~km uncertainty 
in the dimensions of Nix yields a 40\% uncertainty in the volume \citep{kb2019b}. 
The factor of two larger errors in the dimensions of Hydra place correspondingly 
weaker constraints on its volume \citep{weaver2016}.

Within the ensemble of \nbody\ calculations, physical collisions are
exceedingly rare 
(see section 2.4 below). Thus, adopted sizes for the satellites have no impact
on the outcomes of calculations. System stability depends only on adopted 
satellite masses.  With a precise control of satellite masses within each 
calculation, we express results in terms of masses instead of bulk
densities. 
Given the uncertainties in shapes and sizes for each satellite, the 
\nbody\ simulations cannot place direct limits on satellite bulk 
densities; 
these require an error analysis that is independent of the \nbody\ 
simulations as in \citet{kb2019b}.

We define the lifetime of the system $\tau_i$ as the evolution time
between the start of a calculation and the moment when one of the 
satellites is ejected beyond the \pc\ Hill sphere
with an outward velocity that exceeds the local escape velocity, e.g.
$v^2 > 2 ~ G ~ (\mp + \mc) / R_{H, PC}$.  Lifetimes range from 
1--10~yr for very massive (and unlikely) satellite systems to more
than 1~Gyr for systems with the nominal masses.  The uncertainty
of the ejection time is negligible.  When we perform $M$ calculations with 
nearly identical starting conditions, we adopt $\tau_m$ -- the median of 
$M$ different $\tau_i$ -- as the lifetime of the system. For fixed $f$,
the range in $\tau_i$ is $\sim$ a factor of 3--100.  Within the set of 
calculations where we change the mass of only one satellite, we look for 
trends in $\tau_i$ with $f$.

\subsection{Analysis}

To analyze the \nbody\ calculations, we require a formalism to estimate the 
orbital semimajor axis $a$ and eccentricity $e$ of satellites given six phase 
space coordinates.
For a satellite orbiting a single planet with mass $M$, deriving $a$ and $e$ 
is straightforward.  Defining $r$ and $v$ as the instantaneous distance and 
velocity of the satellite, the energy equation
\begin{equation}
\label{eq: akep}
v^2 = GM \left ( \frac{2}{r} ~ - ~ \frac{1}{a_K} \right ) 
\end{equation}
and an equation for the specific relative angular momentum, $h$,
\begin{equation}
\label{eq: ekep}
h^2 = GM ~ (1 ~-~ e_K^2) ~ a_K ~
\end{equation}
yield $a_K$ and $e_K$.  The pericenter $q_K = a_K (1~-~e_K)$ and the apocenter 
$Q_K = a_K (1~+~e_K)$ follow once $a_K$ and $e_K$ are known.

For \nbody\ calculations of satellites orbiting a central binary, a series of 
time steps yields the distance of closest ($R_{min}$) and farthest ($R_{max}$)
distances from the barycenter. With $R_{min}$ and $R_{max}$ as analogs of $q$ 
and $Q$, we derive basic geometric relations for $a$ and $e$ 
\citep[e.g.,][]{sutherland2019}. 
\begin{equation}
\label{eq: ag}
a_g = (R_{max} + R_{min}) / 2 
\end{equation}
and
\begin{equation}
\label{eq: eg}
e_g = (R_{max} - R_{min}) / (R_{max} ~ + ~ R_{min}) ~ .
\end{equation}
These measurements require some care to sample a single circumbinary orbit well 
or to collect a sufficient number of random snapshots over many circumbinary orbits.  
In \citet{kb2019b}, we adopted a similar strategy to identify calculations 
where $\Delta R_g = (R_{max} - R_{min})$ increases with time.

To improve on this approach, \citet{woo2020} developed a fast fourier transform 
(FFT) technique to derive orbital elements based on the restricted three-body 
problem \citep{lee2006}. \citet{lee2006} first define the `guiding center' as 
a reference point in uniform circular motion about the center-of-mass of a binary 
with masses 
\mp\ (primary) and \ms\ (secondary), semimajor axis \abin\ and eccentricity \ebin. 
Within a coordinate system centered on the guiding center, they specify exact 
equations of motion and derive solutions to the linearized problem in terms of 
\rgc, the distance of the guiding center from the barycenter; \efree, the free 
eccentricity of circumbinary particles; $\imath$, the inclination relative to the 
orbital plane of the binary; $t_0$, the time when the guiding center lies on a 
line that connects the two binary components; and several other parameters. For 
the \pc\ binary, where \ebin\ and \efree\ are small, the solutions to the 
linearized equations of motion have a negligible error relative to an `exact' 
solution.

\citet{woo2020} used their FFT technique to derive orbital elements from HST 
observations of the \pc\ satellites \citep{brozovic2015,showalter2015}. Compared 
to orbital fits of the data in standard Keplerian space, the FFT approach accounts 
for the time-variable gravitational potential felt by circumbinary satellites and thus 
yields better estimates for the orbital parameters and their errors. The resulting
orbital elements are almost identical to those inferred from the Keplerian orbital 
fits in \citet{showalter2015}.  Because satellite orbits precess fairly rapidly, 
accurate estimates of $a_{FFT}$ and $e_{FFT}$ require multiple sets of phase space 
coordinates per binary orbit \citep[see also][]{gakis2022}. 

In the \nbody\ calculations described above, we save satellite phase space coordinates 
on time scales of $10^3$--$10^5$~yr over the course of 0.1--3~Gyr. These data are 
insufficient for the \citet{woo2020} algorithm or standard Keplerian fits.  To address 
this problem, \citet{bk2020} developed a fast, approximate method to infer orbital 
elements from a single snapshot of phase space coordinates. They suggest 
two approaches.
The geometric solution is analogous to eqs.~\ref{eq: ag}--\ref{eq: eg}:
\begin{equation}
\label{eq: ageo}
\ageo = [ (R_{max} + R_{min}) ~ - ~ (\Delta R_{+} ~ - ~ \Delta R_{-})] / 2 ~ 
\end{equation}
and
\begin{equation}
\label{eq: egeo}
\egeo = [ (R_{max} - R_{min}) ~ - ~ (\Delta R_{+} ~ - ~ \Delta R_{-})] / (2 \ageo) ~ .
\end{equation}
Here, the $\Delta R_{\pm}$ terms are the extrema of an orbit with \efree\ = 0 in the
\citet{lee2006} formalism. Typically, $\Delta R_{+}$ is larger than $\Delta R_{-}$;
thus \ageo\ is always somewhat smaller than \ag.

Solving the system of equations for the $\Delta$ terms in eq.~\ref{eq: ageo}--\ref{eq: egeo}
requires an iterative technique that converges rapidly. This solution also yields an
approximate \rgc\ and \efree\ in the linearized equations of \citet{lee2006}. For the
four \pc\ satellites, \efree\ and \egeo\ agree very well with $e_{FFT}$ \citep{woo2020}
and the eccentricity derived from the \citet{showalter2015} fit to the HST data
\citep{bk2020}. For \rgc\ and \efree\ estimated from a single epoch in the 
\nbody\ calculation, we adopt single epoch estimates for pericenter and apocenter:
\begin{equation}
\label{eq: qgc}
\qgc\ = \rgc\ ~ (1 ~ - ~ \efree) ~
\end{equation}
and
\begin{equation}
\label{eq: Qgc}
\Qgc\ = \rgc\ ~ (1 ~ + ~ \efree) ~ .
\end{equation}
While these estimates are not the actual pericenter and apocenter that 
would be derived from a well-sampled circumbinary orbit, they provide 
excellent measures of the evolution of orbits during the course of an 
\nbody\ calculation.

\subsection{System Stability}
\label{sec: stability}

From studies of circumstellar and circumbinary planetary systems, 
the four small \pc\ satellites with their nominal masses are 
approximately stable \citep[e.g.,][]{wisdom1980,petit1986,gladman1993,
chambers1996,deck2013,fang2013,fabrycky2014,kratter2014,mahajan2014,
pu2015,morrison2016,obertas2017,weiss2018,sutherland2019}.  Defining the 
mutual Hill radius 
$R_{H, ij} = ((m_i + m_j) / 3 (m_P + m_C))^{1/3} a_i$, where $m_i$ and $m_j$ 
($a_i$ and $a_j$) are the masses (semimajor axes) of a pair of satellites, 
we express the differences in the semimajor axes (e.g., $a_S - a_N$) in 
terms \rhill, $a_i - a_j = K R_{H, ij}$. With this definition, $K_{SN}$ = 
12 for Styx--Nix, $K_{NK}$ = 16 for Nix--Kerberos, and $K_{KH}$ = 10 for 
Kerberos--Hydra.  When orbits are circular and coplanar as in the \pc\ 
satellites, numerical calculations suggest $K \gtrsim$ 8--10 is required 
for stablility. For the nominal masses, the four small satellites are 
barely stable; the Kerberos--Hydra pair is closest to instability.

Aside from the close packing, system stability requires the small 
satellites avoid other pitfalls. All of the satellites are located
close to orbital resonances with the central binary. Test particles
on circular orbits within the 3:1 resonance (near Styx) are unstable on 
short time scales, $\sim$ 100~yr 
\citep[e.g.,][]{ward2006,cheng2014b,bk2015b,giuppone2021}. 
Although test particles within the 4:1 (near Nix), 5:1 (near Kerberos), 
and 6:1 (near Hydra) resonances are stable with no
small satellites in the system, Nix and Hydra together make the 5:1 
resonance much more unstable than the current orbit of Kerberos 
\citep{youdin2012}. Closer to the binary, the 2:1 orbital resonance at 
26.4~\Rp\ (the radius of Pluto, 1~\Rp, is 1188.3~km) lies just inside the
innermost stable circular circumbinary orbit at 28~\Rp\ \citep{holman1999,
doolin2011,kb2019a}. 
Although there are stable orbits much closer to the barycenter
\citep[e.g.,][]{winter2010,giuliatti2013,giuliatti2014,gaslac2019}, 
circumbinary satellites that pass inside the innermost stable orbit are
ejected over 1--10~yr. 

With $K$ = 12, the Styx--Nix pair has some room for oscillations in 
Styx's orbit. At the start of the \nbody\ calculations, Styx (Nix) 
has $e \approx$ 0.001--0.007 (0.002--0.004) from the geometric and
three-body estimates.  When the apocenter of Styx's orbit $Q_S$ and the 
pericenter of Nix's orbit $q_N$ is $q_N - Q_S \approx$ 5000~km, the 
instantaneous K = $q_N - Q_S / r_{H,SN}$ is $\approx$ 10. If Nix maintains 
an eccentricity $e_N \approx$ 0, then Styx has an eccentricity $e_S 
\approx$ 0.023 when the instantaneous $K \approx$ 10. For a limit $K 
\approx$ 8, $e_S \approx$ 0.047. 
When $e_S \gtrsim$ 0.033, Styx crosses the 3:1 resonance near the 
pericenter of its orbit. Resonance crossing will excite the eccentricity
and possibly the inclination. Once $e_S$ reaches 0.14, Styx crosses the
orbit of Nix. If this orbit-crossing excites $e_S$ to 0.22, Styx crosses
the innermost stable orbit and is rapidly ejected from the binary.

For the closer Kerberos--Hydra pair with $K$ = 10, smaller oscillations 
in the orbit of Kerberos lead to instability. These satellites begin the 
calculations with nearly circular orbits, $e_K \approx$ 0.003--0.004 and 
$e_H \approx$ 0.005--0.006. To reach an instantaneous $K$ = 8, the 
eccentricity of Kerberos needs to grow to $e_K \approx$ 0.024. Thus, 
excitations that give Styx--Nix an instantaneous $K \approx$ 10 give 
Kerberos--Hydra $K \approx$ 8. 
Kerberos also lies perilously close to the 5:1 resonance. A 50\% increase
in the initial $e_K$ allows Kerberos to cross the resonance. If
gravitational perturbations from Nix/Hydra and resonance crossing raise
$e_K$ to 0.12 (0.16), Kerberos crosses the orbit of Hydra (Nix). If these
orbit crossings increase $e_K$ to 0.42, Kerberos is inside the innermost 
stable orbit at pericenter and is then rapidly ejected by the central 
binary.

Curiously, the Nix--Hydra pair have a smaller separation in Hill space,
$K_{NH} \approx$ 14, than the Nix--Kerberos pair. With an orbital period 
roughly 1.58 times the orbital period of Nix, Hydra crosses the 3:2 
resonance with Nix at the pericenter of its orbit when 
$e_H \approx$ 0.015.  If oscillations in the orbits of Styx and Kerberos 
excite the orbit of Hydra, then its orbit might also become unstable due 
to this 3:2 resonance. Because Nix is deeper in the potential well, it is 
less likely to be ejected.

Based on these considerations, we anticipate ejections soon after the
eccentricity of either Styx or Kerberos reaches 0.02--0.04. However, Nix 
and Hydra could also excite the inclination of the smaller satellites. 
Although polar orbits are less stable than coplanar orbits 
\citep[e.g.,][]{holman1999,doolin2011,kb2019a}, slightly inclined orbits 
often have longer lifetimes than coplanar orbits. In our analysis, we will 
examine the eccentricities and inclinations of satellites immediately 
preceding each ejection.

\subsection{Examples}

\begin{figure}[t]
\begin{center}
\includegraphics[width=4.5in]{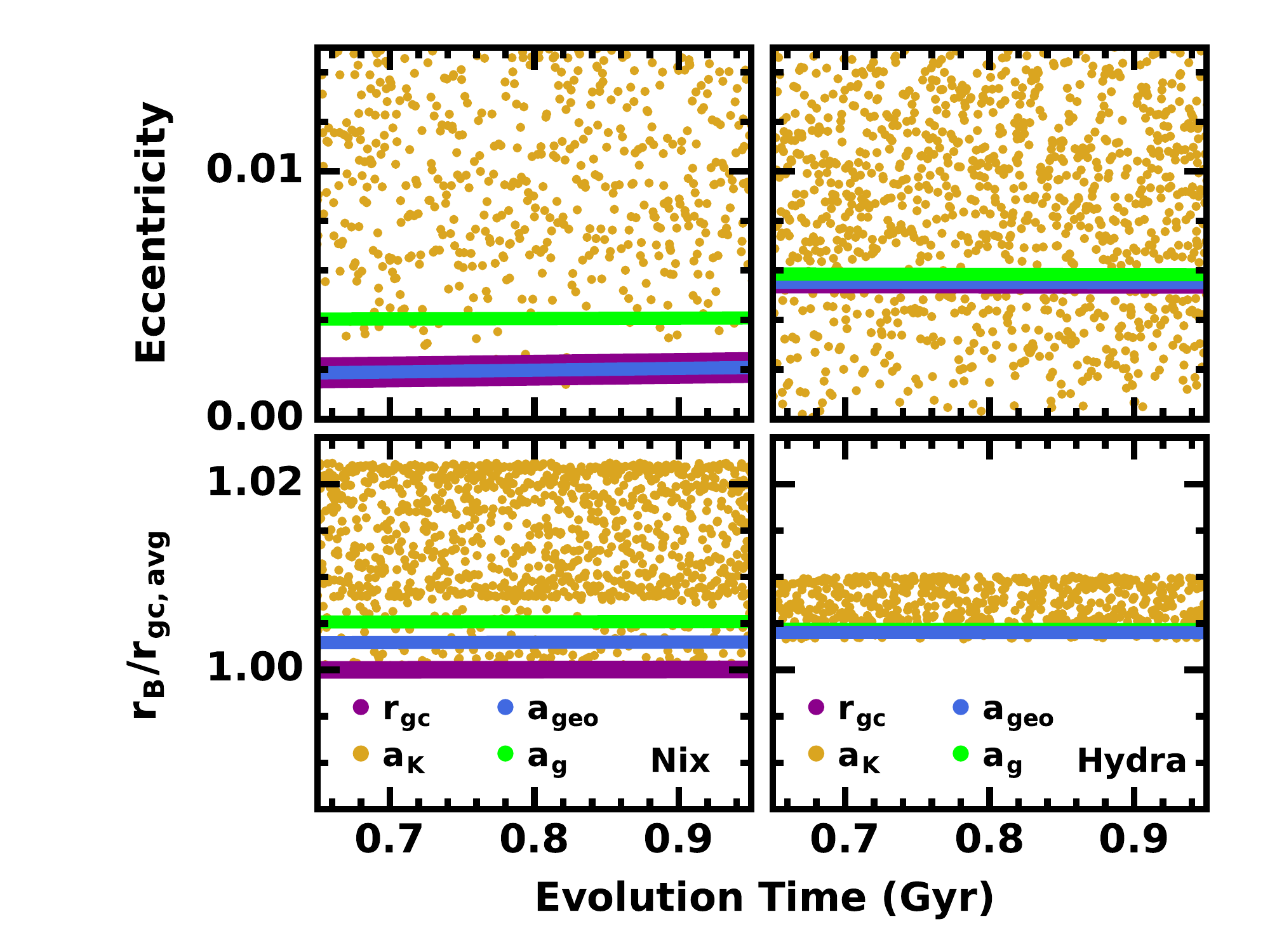}
\vskip -2ex
\caption{
\label{fig: aecomp}
Time evolution of four estimates of the semimajor axis (lower panels)
and the eccentricity (upper panels) for Nix (left panels) and Hydra
(right panels) for a calculations with $f_H$ = 1.1. Other satellites 
have their nominal masses. The legends in the lower panels map colors 
to each 
estimate. The upper panels use the same color scheme as the lower panels.
In each panel, estimates derived from the restricted three-body theory
yield the best results for $a$ and $e$.
}
\end{center}
\end{figure}

To illustrate the application of the \citet{bk2020} formalism, we examine 
a calculation where Styx is ejected at $t \approx$ 1.6~Gyr. The lower 
panels of Fig.~\ref{fig: aecomp} show the evolution of \rgc\ and three estimates 
for the semimajor axis of Nix (left panel) and Hydra (right panel). The
upper panels show the evolution of $e$. 

In the lower panels, the distance of the guiding center from the barycenter
is nearly constant in time. Typical variations over 300~Myr are $\pm$0.001~\Rp.
However, the Keplerian estimate $a_K$ varies wildly with time for all four
satellites. For Nix, the variation in $a_K$ is $\sim$ 2\% of \rgc\ and is 
nearly random. While the randomness persists for Hydra, the fluctuations in
$a_K$ are only 0.5\% of \rgc. Because the potential becomes more spherically
symmetric for orbits with larger \rgc, this trend continues. When the
guiding center radius is $\gtrsim 5 r_{gc,H}$, the orbital energy yields a 
reasonably accurate estimate of $a_K$.

The lower panels of Fig.~\ref{fig: aecomp} also compare the two geometric
estimates of the semimajor axis, \ag\ and \ageo, with \rgc. Because \rgc\ tracks
a circular orbit with \efree\ = 0, \ag\ and \ageo\ are larger than \rgc\ when
\efree\ is larger than zero. In the Nix example, the orbital excursions due
to the binary potential (i.e., $\Delta R_+$ and $\Delta R_-$) are large,
$\sim$ 0.01--0.03~\Rp. Taking these excursions into account places \ageo\ inside
\ag. Moving outward in the system to Hydra, $\Delta R_+$ and $\Delta R_-$ are
smaller, $\lesssim$ 0.005~\Rp. While \ageo\ still lies inside \ag, the
difference is not obvious in the Figure.

The upper panels of Fig.~\ref{fig: aecomp} demonstrate that \efree\ and
\egeo\ yield similar results for the eccentricity of Nix's orbit. 
Measurements of \efree\ are variable at the $\pm$0.0004 level due to the 
accuracy limitation of the estimator \citep[see Fig. 3 of][]{bk2021};
\egeo\ is nearly constant in time. 
The \eg\ values are a
factor of two larger and also roughly constant in time. Once again the
Keplerian estimate $e_K$ is nearly random; even the median of $e_K$ provides
a poor measure of the eccentricity. 

For the orbit of Hydra, three estimates -- \eg, \egeo, and \efree\ -- are
essentially identical. Estimates based on the restricted three body model,
\egeo\ and \efree, are indistinguishable and vary little in time. The
basic geometric value is somewhat larger. Although the Keplerian approach
yields a more accurate semimajor axis for Hydra than for the other satellites, 
the eccentricity measure is not useful. 

As with the semimajor axis, the Keplerian eccentricity is more accurate for 
orbits more distant from the barycenter. Because $e$ is more sensitive to 
the binary potential than $a$, we recommend using the energy and angular 
momentum equations for $a$ and $e$ only for orbits with $a_K \gtrsim$ 600~\Rp. 
Between 100~\Rp\ and 600~\Rp, \ag\ and \eg\ are fairly reliable. Inside of
100~\Rp, we prefer the three-body estimates.

\begin{figure}[t]
\begin{center}
\includegraphics[width=4.5in]{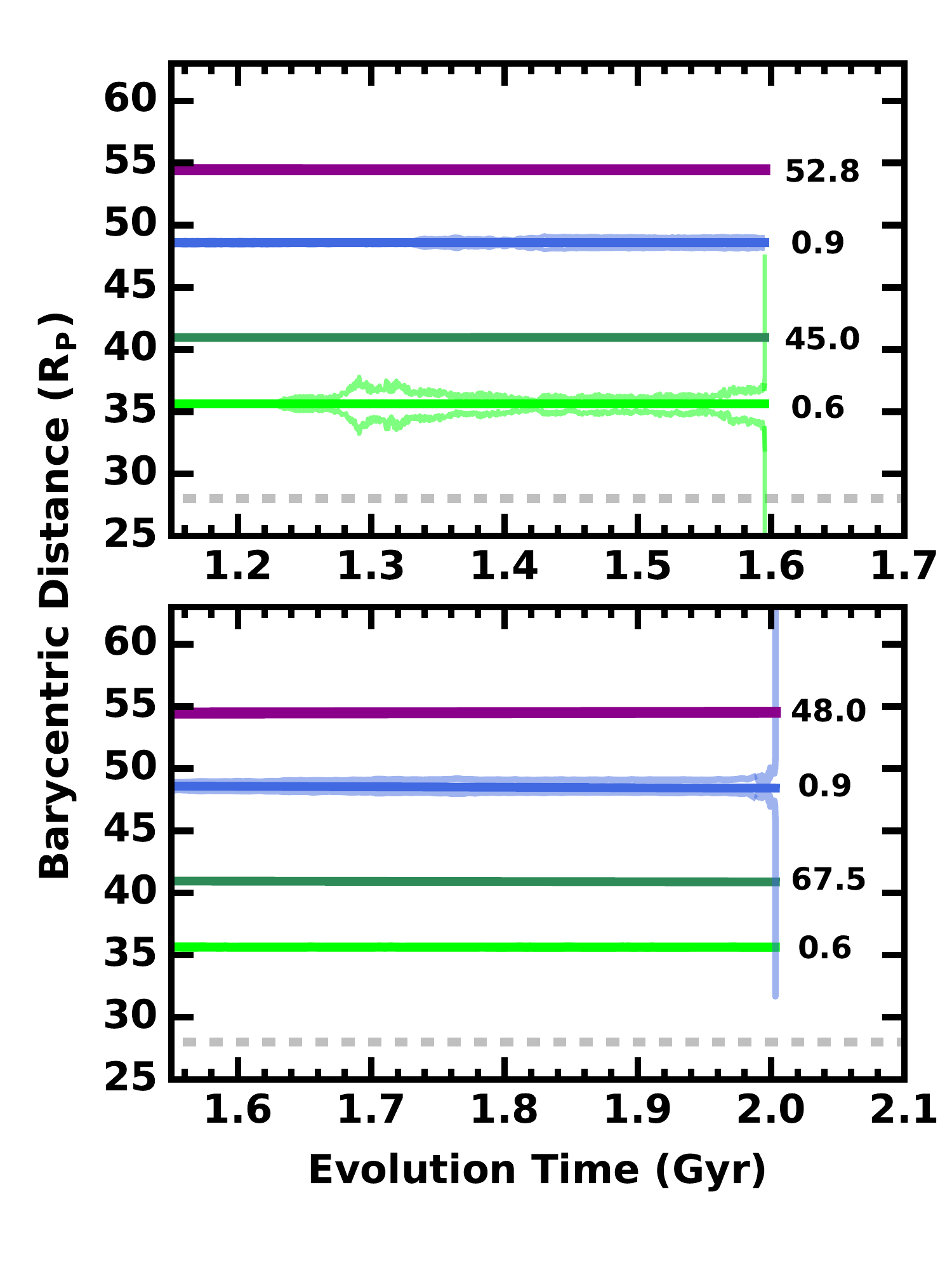}
\vskip -4ex
\caption{
\label{fig: nhtime}
Time evolution of the barycentric distance for Styx (light green),
Nix (dark green), Kerberos (blue), and Hydra (purple). 
The very light green (blue) curve plots $q_S$ and $Q_S$ 
($q_K$ and $Q_K$) as defined in the text.
The horizontal dashed grey line near the bottom of each panel indicates 
the location of the innermost stable distance for circular orbits 
in the orbital plane of the central binary \citep{kb2019a}.
The masses (in units of $10^{18}$~g) for each satellite in the 
simulation are listed to the right of each curve.  
}
\end{center}
\end{figure}

Fig.~\ref{fig: nhtime} illustrates the time evolution of the barycentric 
distance of the four satellites. 
Lighter curves in each panel plot $q_i$ and $Q_i$.
In every calculation, the more massive Nix and Hydra perturb the orbits of 
Styx and Kerberos. Sometimes the perturbations are small; variations in the 
barycentric distances of Styx and Kerberos are roughly constant in time.  
In other cases, oscillations \rgc\, $q$, and $Q$ steadily grow with time. 
When the combined mass of the satellite system is several times the nominal 
combined mass, orbits evolve rapidly; Styx or Kerberos or both are ejected. 
Systems with a mass closer to the nominal combined mass evolve very slowly. 
However, the excitations eventually cross a threshold, becoming more dramatic 
with every orbit.  At least one of the small satellites is then ejected from the system.

In the top panel of Fig.~\ref{fig: nhtime}, Hydra has a mass 10\% larger than 
its nominal mass.  Other satellites have their nominal masses. During the first 
Gyr of evolution, oscillations in 
$q$ and $Q$
for Styx and Kerberos are fairly stable.  At $\sim$ 1.25~Gyr, oscillations in 
the orbit of Styx become more obvious. 
Although it seems likely at this point that Styx will soon cross the orbit of Nix,
the orbit gradually recovers and returns to a low $e$ for the next 250~Myr. 
During this period, the orbit of Kerberos develops a modest eccentricity. 
Eventually, Styx crosses the orbits of Nix and Kerberos and ventures well 
inside the unstable region. The central binary then ejects it from the system.

The lower panel of Fig.~\ref{fig: nhtime} shows an example where the mass of Nix 
is 50\% larger than its nominal mass. Other satellites have their nominal masses. 
For roughly 1.5~Gyr, oscillations in the orbit of Kerberos are steady. In response 
to the larger mass of Nix, the orbit of Styx moves slightly closer to the system 
barycenter.  Oscillations in this orbit are modest and also remain roughly constant 
in time.  At $\sim$ 2~Gyr, oscillations in Kerberos' orbit grow rapidly. Unlike the 
example in the top panel of Fig.~\ref{fig: nhtime}, Kerberos does not excite 
fluctuations in Styx's orbit.  The oscillations in Kerberos' orbit simply grow until 
Kerberos crosses the orbits of Nix and Hydra. Gravitational kicks from Nix and Hydra
eventually push Kerberos well inside the unstable region surrounding the central binary, 
which ejects it from the system.

\begin{figure}[t]
\begin{center}
\includegraphics[width=4.5in]{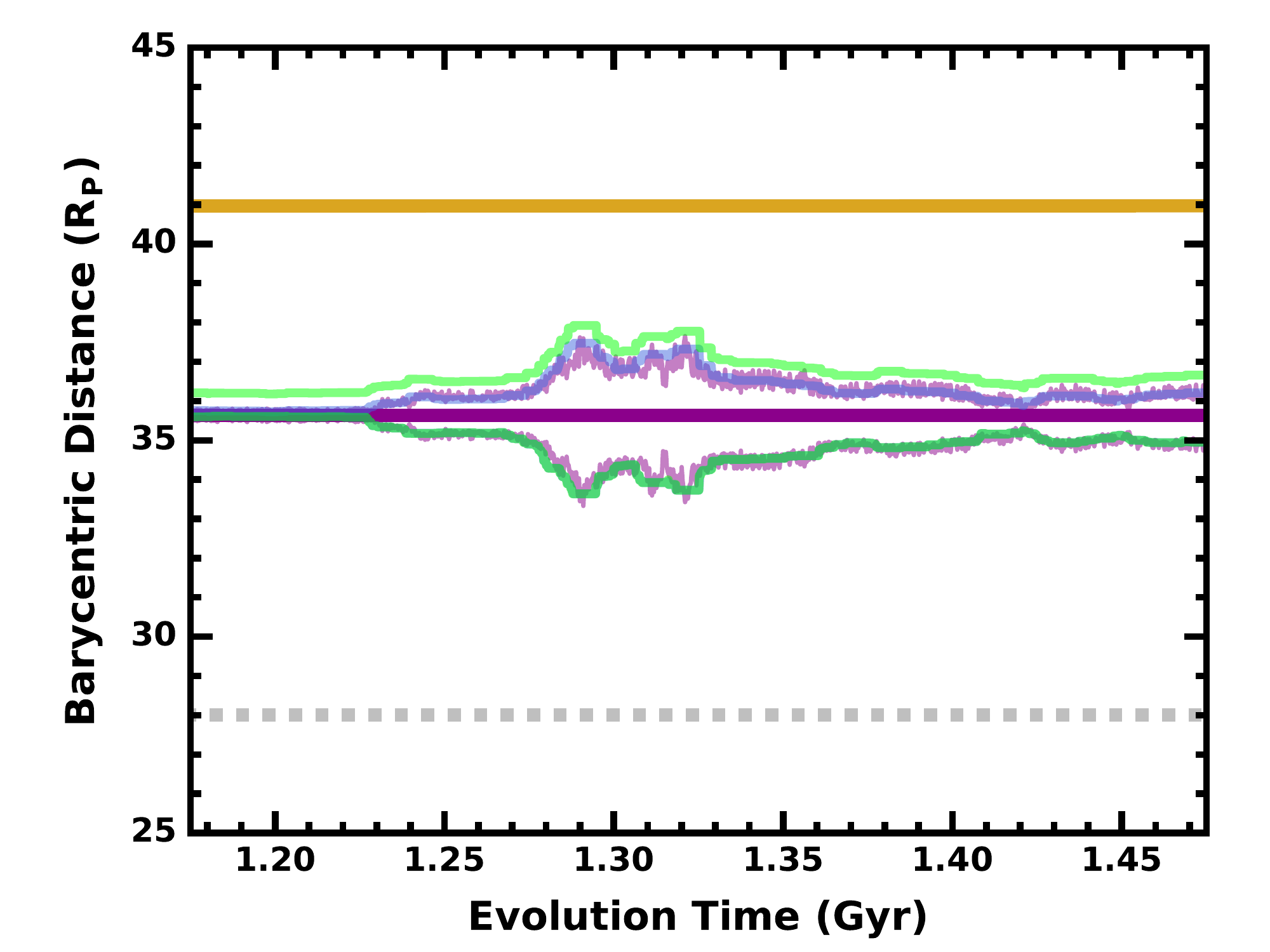}
\vskip -2ex
\caption{
\label{fig: styx}
Expanded view of the time evolution of the barycentric distance for 
Styx and Nix from the top panel of Fig.~\ref{fig: nhtime}, where 
$f_H$ = 1.1. Other satellites have their nominal masses. The light tan
curve shows the orbit of Nix. For the orbit of Styx, the dark purple 
curve plots \rgc; $q$ and $Q$ are shown for the basic geometric model 
(light green), the geometric solution from the restricted three-body 
problem (blue), and for the model for \efree\ (light purple).
The basic geometric values have a larger excursion in the orbit than 
values derived from the restricted three-body problem.
}
\end{center}
\end{figure}

Fig.~\ref{fig: styx} illustrates the evolution of $q_S$ and $Q_S$ in an expanded
version of the top panel of Fig~\ref{fig: nhtime}. Throughout this part of the 
evolution, \rgc\ (dark purple curve) is nearly constant in time. At $\sim$ 1.23~Gyr,
$q_{gc}$ and $Q_{gc}$ (light purple curves) begin to grow, each reaching $\sim$ 
2~\Rp\ from \rgc\ (\efree\ $\approx$ 0.06). After $\sim$ 50~Myr, \qgc\ and \Qgc\
begin a long decline and then undergo and smaller oscillation on a longer times scale.
During the main oscillation in \qgc\ and \Qgc, \rgc\ moves slightly inward and then
returns to its original value.

\begin{figure}[t]
\begin{center}
\includegraphics[width=4.5in]{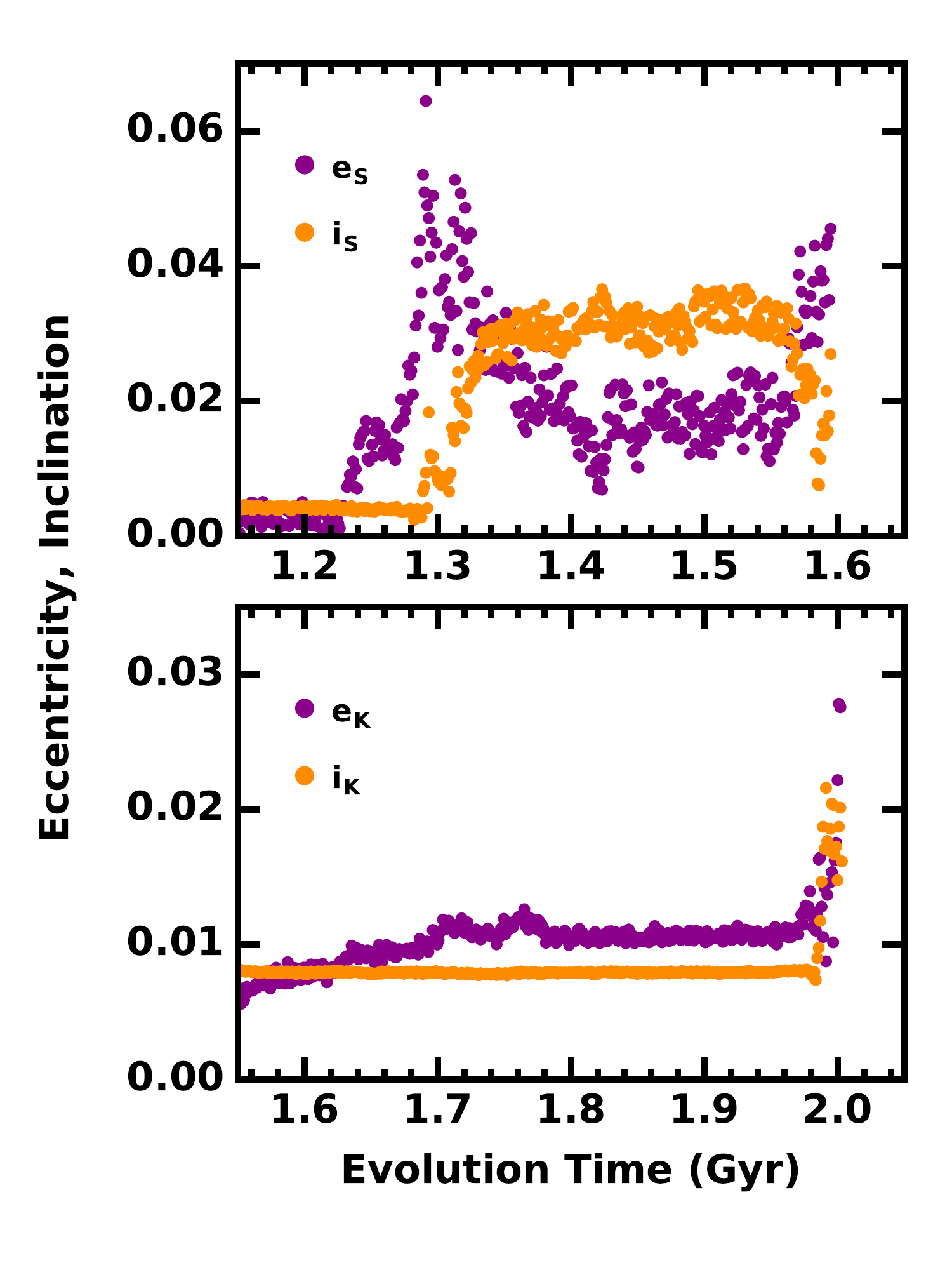}
\vskip -5ex
\caption{
\label{fig: eitime}
Time variation of \efree\ and $\imath$ for Kerberos (lower panel, from the
calculation shown in the lower panel of Fig.~\ref{fig: nhtime}) and 
Styx (upper panel, from the calculation shown in the upper panel of 
Fig.~\ref{fig: nhtime}). In the lower panel, the orbital inclination of
Kerberos is approximately constant for nearly 2~Gyr and then rises abruptly
just before Kerberos is ejected. In the upper panel, the inclination of 
Styx begins to rise when $e$ reaches a maximum level and then finds a 
plateau as $e$ returns to a stable value. Once $e$ rises again, $\imath$ 
drops; Styx is then ejected.
}
\end{center}
\end{figure}

Other curves in this Figure show the variation of \qgeo\ and \Qgeo\ (light blue curves)
and \qg\ and Qg\ (light green curves). Throughout this time sequence, pericenter
and apocenter derived from the basic geometric model have larger fluctuations than
those estimated from the restricted three-body model. The \ageo\ and \Qgeo\ values
track the guiding center values very closely and have somewhat smaller fluctuations.

All of these estimates yield a much better characterization of the orbit than the
Keplerian estimates derived from eqs.~\ref{eq: akep}--\ref{eq: ekep}. With $a_K$
and $e_K$, the pericenter of Styx' orbit sometimes passes close to or inside the
unstable region surrounding the binary; at these times, apocenter passes dangerously
close to Nix. This behavior is a function of the non-spherical potential of the
central binary, which generates small radial excursions in the orbit and can `fool'
the Keplerian estimators into thinking Styx is much closer to Nix than it actually
is. 

Although the basic and more advanced geometric estimates track the orbit well, 
it is important to take some care in setting the appropriate window to derive 
$R_{min}$ and $R_{max}$ from a time sequence within an \nbody\ calculation
\citep[see also][]{sutherland2019}. Here, we used a time-centered approach,
inferring $R_{min}$ and $R_{max}$ from 100 samples before and another 100 samples
after a given time $t$. While fewer samples enable accurate estimates of \ag\ or 
\ageo, they also resulted in smaller differences between $q$ and $Q$. Among the
set of calculations analyzed for this paper, larger samples did not change the
results significantly.

To understand the behavior in the evolution of Styx and Kerberos in 
Fig.\ref{fig: nhtime}, we consider the variation in $e$ and $\imath$
(Fig.~\ref{fig: eitime}). In the calculation where Kerberos is ejected 
(lower panel),
$e$ slowly rises from $\sim$ 0.005 at 1.5~Gyr to 0.011 at 1.8~Gyr,
remains roughly constant for $\sim$ 170~Myr, and then rises dramatically
as Kerberos is ejected (Fig.~\ref{fig: eitime}). In contrast, $\imath$ 
is nearly constant at 0.0078 for 1.98~Gyr, drops somewhat as $e$ begins 
to grow, and then follows the dramatic increase in $e$ as Kerberos crosses
the orbits of Nix and Hydra, ventures into the unstable zone close to \pc,
and is then ejected. 

The evolution of Styx in the upper panel of Fig.~\ref{fig: eitime} is more
interesting. Well before ejection, the orbital eccentricity of Styx first
grows dramatically and then varies between 0.03 and 0.06 at 1.22--1.32~Gyr.
During this period, Styx has not crossed the orbit of Nix or the innermost
circumbinary orbit, but a much larger growth in $e$ would place Styx in peril.
Just after $e$ reaches a maximum, $\imath$ begins to rise and reaches a
maximum of $\imath \approx$ 0.03 at $\sim$ 1.34~Gyr. During this period $e$
drops to 0.01--0.02 and maintains this level for close to 200~Myr. Eventually,
the eccentricity begins to grow (and the inclination drops). Styx is then 
ejected.

Among the $\sim$ 300 calculations where the combined mass of the 
satellites is no more than twice the total nominal mass, all follow 
one of the two paths summarized in Fig.~\ref{fig: eitime}. In most 
systems, $e_S$ or $e_K$ slowly rises with little or no change in 
$\imath_S$ or $\imath_K$. Eventually, the eccentricity of one satellites 
begins to grow dramatically; $\imath$ then rises to similar levels and 
the satellite is ejected. In other calculations, a satellite maintains a 
large $\imath$ for a long period before an ejection. In these cases, the 
growth of $\imath$ allows the satellite to avoid Nix and Hydra, 
stabilizing the satellite's orbit for many Myr. Eventually, perturbations
from Nix and Hydra induce the satellite to cross into the unstable region 
close to the binary and the satellite is ejected.

\subsection{Collision Rates}
\label{sec: coll}

In all calculations where the \nbody\ code resolves collisions, small
satellites are ejected before they collide with one another. Combining
the examples in the previous section with the orbital architecture, 
we demonstrate that physical collisions are unlikely. We approximate 
the orbits of Styx/Kerberos (Nix/Hydra) as coplanar ellipses (circles) 
relative to the system barycenter. Collisions only occur near the 
apocenters of the orbits of Styx (for collisions with Nix) and Kerberos 
(for collisions with Hydra). Defining a time period $\Delta t$ when a
collision is possible, the probability a satellite with orbital period
$P_i$ occupies that part of its orbit is $\Delta t / P_i$. The other 
satellite with orbital period $P_j$ occupies this region once per orbit. 
Thus, the collision probability is 
\begin{equation}
\label{eq: pcoll}
p_{coll} = \frac{2 \Delta t}{P_i P_j} ~ ,
\end{equation}
where the factor of two results from two crossings per orbital period $P$.

For a quantitative estimate of $\Delta t$, we consider the equation 
of the barycentric distance of an elliptical orbit for Styx or Kerberos, 
$r_i = a_i (1 - e_i^2) / (1 + e ~ {\rm cos} ~ \theta_i)$. Here, 
the angle  $\theta_i$ is measured counterclockwise from the $x$-axis 
in the orbital plane. Defining 
$R_{tot} = R_i + R_j$ as the sum of the physical radii, 
we seek solutions for $\theta_i$ where the small satellite has a 
distance from the barycenter
\begin{equation}
\label{eq: dcoll}
a_j - R_{tot} \le r_i \le a_j + R_{tot}
\end{equation}
For an adopted $e_i$, it is
straightforward to solve for the two angles that define the minimum,
$\theta_{i,min}$, and maximum, $\theta_{i,max}$ distances where collisions 
can occur. The interaction time $\Delta t_i$ follows from 
$\Delta \theta_i = \theta_{i, max} - \theta_{i, min}$ and 
application of Kepler's second law. 

These solutions suggest the typical interaction time $\Delta t$ is small. 
When Styx has $e_S \le$ 0.1409, collisions cannot occur. As $e_S$ grows 
from 0.141 to 0.1422, $\Delta t$ rises from $\sim$ 50~sec to $\sim$ 
1650 sec. For larger $e_S$, the interaction time is a slowly decreasing 
function of $e_S$. When
Kerberos has $e_K \le$ 0.1199, collisions with Hydra are impossible. This
system has a maximum $\Delta t \approx$ 2500~sec for $e_K$ = 0.1209 and
a slowly decreasing $\Delta t$ at larger $e_K$. In both cases, the maximum
interaction time is at apocenter. The corresponding collision probabilities
are $p_{coll} \approx 0.0019$ per Nix orbit for collisions with Styx and
$p_{coll} \approx 0.0018$ per Hydra orbit for collisions with Kerberos.
The time scale from orbit crossing to ejection is typically a few orbits
of Nix or Hydra. Thus, the likelihood of a physical collision is small.

Other approaches to derive $\Delta t$ yield similar results \citep[e.g.,][]{opik1951,
wetherill1967,
rickman2014,jeongahn2017}. All express the probability of a collision in a form similar
to eq.~\ref{eq: pcoll}. Our method has the advantage of avoiding linear
or parabolic approximations to the trajectories for analytical solutions
and the intricacies of more involved numerical estimates. Although 
we ignore the exact shape of a circumbinary orbit, corrections due
to the circumbinary potential should be small.

Generalizing the method to an elliptical orbit for the more massive 
target and an inclined orbit for the impactor is straightforward. 
Elliptical orbits for Nix/Hydra do not change the probabilities 
significantly. At apocenter, Styx (Kerberos) lies at a height $z_S$ 
($z_K$) above the orbital plane of Nix (Hydra), where 
$z_i = a_i ~ {\rm sin} (\imath_i - \imath_j)$. Requiring 
$|z_i| \le R_{tot}$ when orbits cross implies 
$\Delta \imath = |\imath_i - \imath_j| \le$ 0.001 (0.00077) for 
Styx--Nix (Kerberos--Hydra) collisions. At the start of each calculation
$\Delta \imath \approx$ 0.0029 for Styx--Nix and 0.0028 for 
Kerberos--Hydra. In all of the calculations, the inclination difference
between Styx--Nix and Kerberos--Hydra grows with time. Thus, physical
collisions among the small satellites are impossible.

Despite the lack of physical collisions, orbit crossings often place
Styx (Kerberos) within the Hill sphere of Nix (Hydra). Replacing 
$R_{tot}$ in eq.~\ref{eq: dcoll} with the radius of the Nix or Hydra 
Hill sphere results in factor of ten larger probabilities for strong
dynamical interactions instead of physical collisions when Styx or
Kerberos are at apocenter. The large satellite inclinations 
$\Delta \imath \approx$ 0.01 also place Styx (Kerberos) on the edge 
of the Nix (Hydra) Hill sphere during orbit crossings. Entering the Hill
spheres of Nix and Hydra provide additional perturbations to the orbits
of Styx and Kerberos.


In the next section, we discuss results from the ensemble of \nbody\ 
calculations. After examining the ejection frequency for Styx and Kerberos
as a function of initial system mass, we derive the frequency of systems 
where $\imath$ maintains an elevated level prior to ejection. We conclude
with new estimates of the total system mass and illustrate the impact of
the masses of Styx and Kerberos on the ejection time scale.

\section{Results} \label{sec: results}

Once a calculation begins, all of the evolutionary sequences follow the 
same trend. After a period of relative stability where the orbital 
parameters of the system are roughly constant in time, the motion of at 
least one satellite begins to deviate from its original orbit.  These 
deviations grow larger and larger until the orbits cross
the 3:1 (Styx) or the 5:1 (Kerberos) resonance with the central binary.
After resonance crossing, orbits are excited to larger $e$ and $\imath$.
Although Styx and Kerberos can maintain a state of higher $e$ and $\imath$
for awhile, resonance crossing and perturbations by Nix and Hydra 
eventually lead to orbit-crossings.
After orbits begin to cross, the motion of either Styx or Kerberos 
rapidly carries it inside the region close to \pc\ where circumbinary 
orbits are unstable. After a few passes, 
at least one satellite -- usually Styx or Kerberos -- is ejected from the 
system.

\subsection{Ejection Statistics}
\label{sec: res-ej}

To examine ejection statistics for the ensemble of \nbody\ calculations, we
begin with a brief summary of the model parameters. In one set of calculations,
we vary the mass of Nix or Hydra and set other satellites at their nominal masses.
As summarized in Table~\ref{tab: ejfrac}, we have 38 (50) completed calculations
for $f_N > 1$ ($f_H >$ 1). The character of the evolution changes when Hydra or
Nix are more than twice their nominal masses. Thus, we consider two sets of results
for each model sequence.

In the calculations where the masses of all satellites are multiplied by the same
factor $f$, we have sets with $f_s = f_K$ = 1.0, 1.5, 2.0, and 3. For $f \ge$ 1.5,
nearly all calculations are complete. While most with $f$ = 1.25 are complete,
only 10 with $f$ = 1 have finished. Comparisons among the sets with $f \ge 1.5$
indicates that ejection frequencies are independent of $f_S$ and $f_K$. Thus, we
combine statistics into one row of Table~\ref{tab: ejfrac}. For simplicity, we
follow this procedure for sets with $f$ = 1.00 and 1.25.

\begin{figure}[t]
\begin{center}
\includegraphics[width=4.5in]{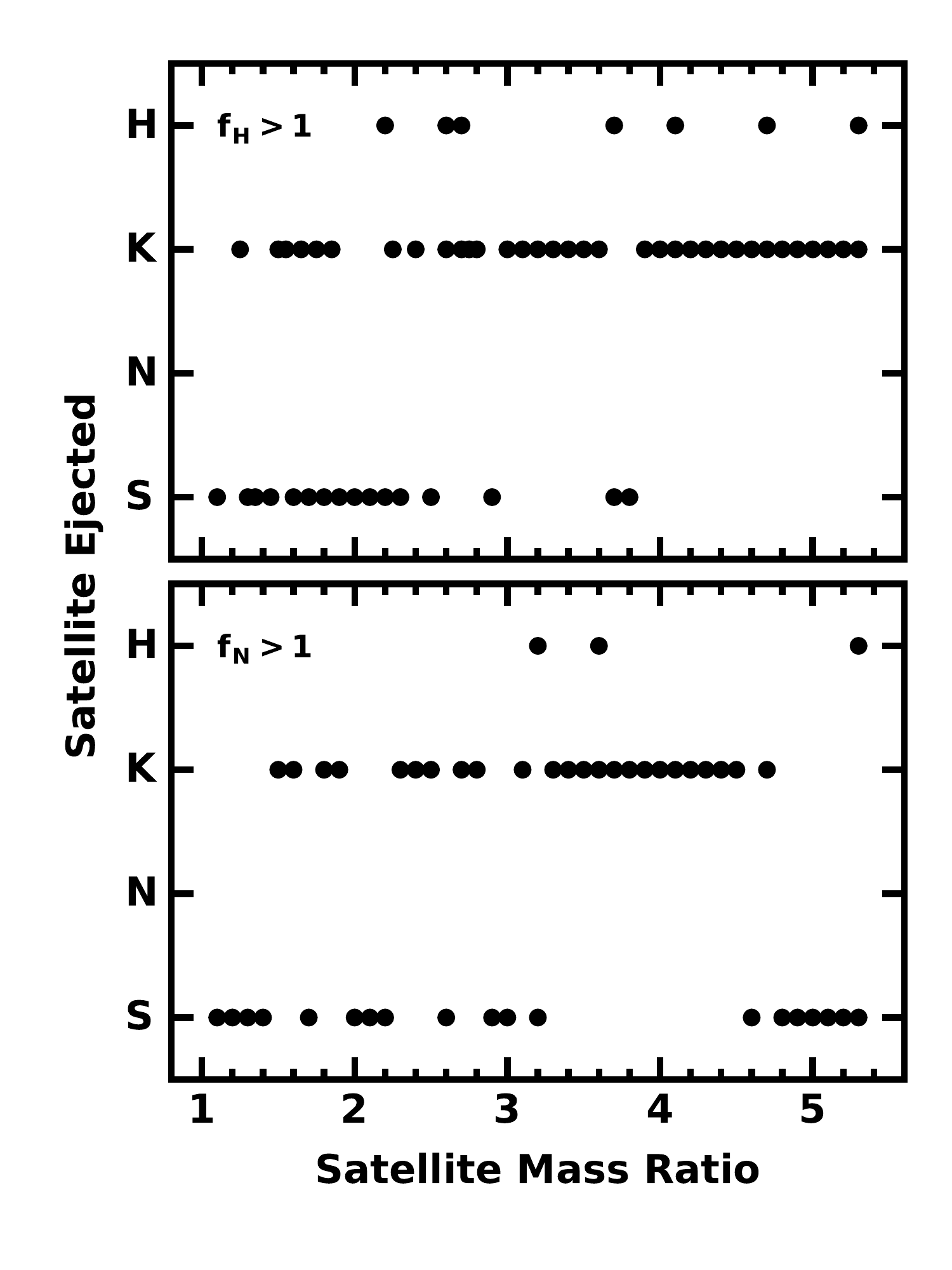}
\vskip -5ex
\caption{
\label{fig: eject}
Ejection diagram for calculations with $f_N > 1$ (lower panel) and
$f_H > 1$; other satellites have their nominal masses. Letters on the
y-axis correspond to the satellites, with `S' for Styx, `N' for Nix,
`K' for Kerberos, and 'H' for Hydra. Nix is never 
ejected. Hydra ejections only occur in very massive systems. Kerberos
is usually ejected when Hydra is massive. A massive Nix ejects either
Styx or Kerberos roughly half of the time.
}
\end{center}
\end{figure}

Figure~\ref{fig: eject} illustrates results for the full set of calculations with
$f_N > 1$ (lower panel) and $f_H > 1$ (upper panel). At the largest masses in the
lower panel, Kerberos is rarely ejected. Among systems with slightly smaller masses,
$f_N$ = 3.0--4.5, Kerberos is almost always ejected. At still lower masses, Styx
and Kerberos have similar ejection probabilities. Styx ejections dominate the lowest
masses.  Kerberos ejections dominate the set of calculations with $f_H > 1$ in the
upper panel. Among the more massive systems with $f_H > 2$, Styx ejections are rare. 
Below this limit, Styx ejections dominate. 

Surprisingly, ejections of Styx or Kerberos are sometimes accompanied by an ejection
of Hydra. Nix is never ejected. Of the ten calculations where Hydra is ejected, Styx
accompanies Hydra out of the \pc\ system 4 times. All of these systems have massive
satellites with $f_N > 2$ or $f_H > 2$ and follow a common pattern. Once Styx 
(Kerberos) begins to cross the orbit of Nix (Hydra), Hydra's orbit begins a small
oscillation, crosses the 3:2 resonance with Nix at pericenter, develops a much larger
oscillation in its orbit, and is then ejected.

To supplement Fig.~\ref{fig: eject}, the first four rows of Table~\ref{tab: ejfrac}
summarize the frequency of Styx, Kerberos, and Hydra ejections in this set of calculations.
The dominance of Kerberos (Styx) ejections at large (small) systems masses has a simple
physical explanation in Hill space. In massive systems with $f_H > 2$ or $f_N > 2$, 
the satellite with the smallest $K$ (Kerberos) is most prone to develop oscillations that
lead to orbit crossings. Lower mass systems with $f_H < 2$ or $f_N < 2$ always have
$K > 8$ and thus meet the minimum criterion for stability. Here, Styx's lower mass and
proximity to the innermost stable orbit make it a better candidate for ejection.

\begin{deluxetable}{ccccccc}
\tablecolumns{7}
\tablewidth{15cm}
\tabletypesize{\normalsize}
\tablenum{2}
\tablecaption{Ejection Fractions}
\tablehead{
\colhead{Model} &
\colhead{~~~N~~~~} &
\colhead{~Styx~~} &
\colhead{~~~Nix~~~~} &
\colhead{~Kerberos~~} &
\colhead{~Hydra~~}
}
\label{tab: ejfrac}
\startdata
$f_N > 2$   & 29 & 0.34 & 0.00 & 0.66 & 0.10 \\
$f_N \le 2$ & 9  & 0.67 & 0.00 & 0.33 & 0.00 \\
$f_H > 2$   & 35 & 0.20 & 0.00 & 0.80 & 0.20 \\
$f_H \le 2$ & 15 & 0.53 & 0.00 & 0.47 & 0.00 \\
$f$ = 1.00  & 10 & 0.80 & 0.00 & 0.20 & 0.00 \\
$f$ = 1.25  & 44 & 0.24 & 0.00 & 0.76 & 0.00 \\
$f$ = 1.50  & 63 & 0.32 & 0.00 & 0.68 & 0.00 \\
$f$ = 2.00  & 67 & 0.28 & 0.00 & 0.73 & 0.01 \\
\enddata
\end{deluxetable}

These considerations hold in the larger set of calculations with a fixed $f$ for all
four satellites (Table~\ref{tab: ejfrac}, rows 5--8). In massive systems ($f$ = 1.25,
1.5 and 2), Kerberos is usually ejected. Styx dominates ejections for $f$ = 1, but the
sample size is smaller. The lone calculation with a Hydra ejection occurs when $f$ = 2.
Another $f$ = 2 calculation is the only one where Styx and Kerberos are ejected. 

\subsection{Signals}
\label{sec: sec-signal}

Fig.~\ref{fig: eitime} illustrates two classes of evolutionary sequences in the
\nbody\ calculations. Often, the eccentricity of Styx or Kerberos gradually increases
until the satellite starts to cross the orbit of one of the massive satellites. The
satellite is then scattered across the innermost stable orbit, where \pc\ eject it
from the system. In other sequences, $e$ grows but stops short of orbit crossing due
to a substantial increase in the inclination which reduces $e$. The system remains in
this state for awhile before the high inclination satellite begins another foray at
high $e$. This time, the satellite is ejected.

To understand the frequency of the two different types of sequences, we examine the
evolutionary history of each \nbody\ calculation with $f \le 1.5$, $f_N \le 2$, or
$f_H \le$ 2. For each of the 140 calculations in this sample, we derive the average
$e$ and $\imath$ during the first 5\% to 10\% of the sequence. This exercise yields
$(e_i, \imath_i)$ = (0.0073, 0.0044) for Styx and (0.0042, 0.0077) for Kerberos. For
the satellite to be ejected at the end of the time sequence, we then search for the 
first time $t_0$ where $\imath$ rises above 0.01 after $e$ exceeds 0.01.  With $t_0$ 
established, we verify that $e$ and $\imath$ are more than 10$\sigma$ larger than the 
average values and that both remain larger than 0.01 until a satellite is ejected. 
We define the fractional delay in an ejection as $D = (t_0 - t_f) / t_f$, where $t_f$ 
is the time the satellite leaves the \pc\ system. 

In addition to identifying $t_0$ and $D$, we investigate the maximum $e$ and $\imath$ 
for ejected satellites prior to ejection. In systems with small $D$, the maximum
inclination for Styx is 0.01--0.02. When $D$ is large, Styx's $\imath$ also tends 
larger, 0.05--0.10. Several time steps before ejection, $e \approx \imath$; in these
cases, Styx follows an evolution similar to that in Fig.~\ref{fig: eitime} where an
increase in $e$ generates a rise in $\imath$. For Kerberos, there is little correlation
between $D$ and values of $e$ and $\imath$ before ejection. 

\begin{figure}[t]
\begin{center}
\includegraphics[width=4.5in]{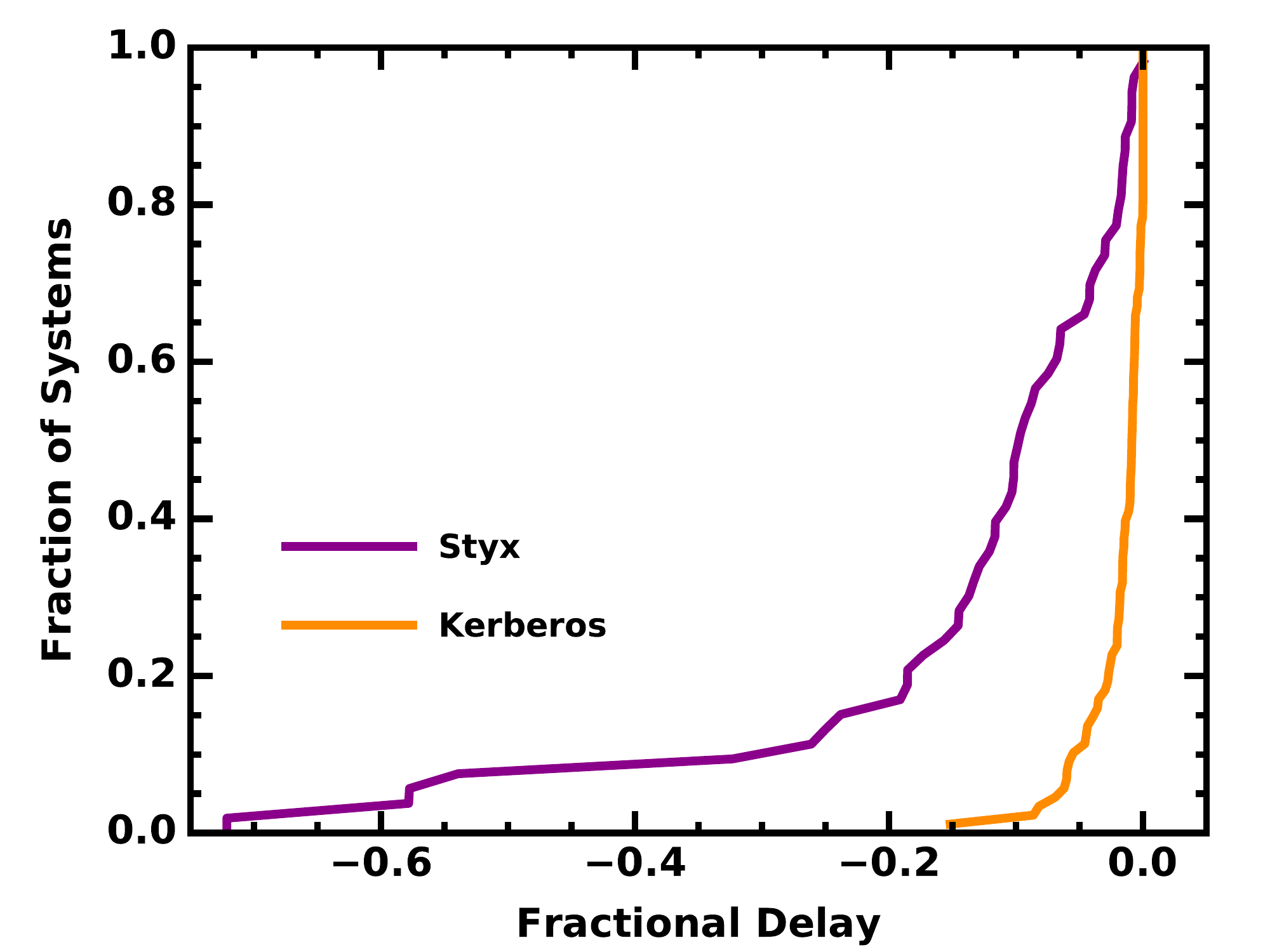}
\caption{
\label{fig: idelay}
Cumulative distribution of the fractional time delay between
the time when $\imath$ first rises above 0.01 and the time of
ejection.  In systems where Kerberos is ejected (orange curve), 
$\imath$ typically becomes larger than 0.01 just prior to
ejection. When Styx is ejected, $\imath$ is often large well
before ejection.
}
\end{center}
\end{figure}

Fig.~\ref{fig: idelay} shows the cumulative distribution of time delays for Styx
(purple curve) and Kerberos (orange curve). Kerberos rarely produces a significant
`signal' that it will eventually be ejected. In $\sim$ 20\% of the ejections, Kerberos 
does not signal at all: $\imath$ rises above 0.01 in the time step immediately prceeding 
the ejection. In another 37\%, the delay is extremely short with $t_0 \gtrsim 0.99 ~ t_f$. 
In the remaining sequences, the delay is 1\% to 10\% of the system lifetime; the longest
delay is 15\%.

In contrast, Styx almost always provides a robust signal for its impending ejection. In
only 10\% of the ejections, Styx's signal is weak with $t_0 \gtrsim$ 0.99 $t_f$. Roughly
half of the calculations find Styx with $\imath \gtrsim$ 0.01 at a time 10\% or more before
$t_f$. In some remarkable cases, Styx has a large $e$ and $\imath$ when $t_0 \lesssim 0.5 ~ t_f$.
Somehow, these systems stay on the edge of instability for extended periods of time before
an ejection.

In nearly all of these examples, the low mass satellite that is not ejected maintains much
smaller $e$ and $\imath$ throughout the period where the other satellite is on the verge of
ejection. Only one calculation has an ejection of Kerberos and Styx. Among the set of
calculations with an ejection of Kerberos, Styx never has time to react to the orbital 
gyrations of Kerberos. After the orbit of Kerberos is perturbed, it rushes to an ejection.
When Styx is to be ejected, Kerberos often reacts slightly to Styx's oscillations in $e$
and $\imath$. Curiously, these never lead to ejection. As Styx leaves the system, Kerberos
maintains a stable orbit.

The difference between Styx and Kerberos in Fig.~\ref{fig: idelay} is probably a function 
of orbital separation in Hill units, $K$. As the satellite with the smallest $K$ and flanked
by two massive satellites, Kerberos is closest to ejection at the start of each calculation. 
Small kicks from either Nix or Hydra are sufficient to increase $e$ and $\imath$ above 0.01.
When that occurs, ejection rapidly follows. As discussed above, Styx has more room for its
orbit to fluctuate. Despite its proximity to the innermost stable orbit, it can more easily
trade off $\imath$ for $e$ and remain on the edge of ejection for many years. 

\subsection{Constraints on the System Mass}

In \citet{kb2019b}, we considered the evolution of `heavy' satellite 
systems,
where Nix and Hydra have the masses listed in Table~\ref{tab: init} and 
the masses of Styx ($4.5 \times 10^{18}$~g) and Kerberos ($1.65 \times 10^{19}$~g)
are consistent with those reported in \citet{brozovic2015}. With lifetimes
$\tau_i \approx$ 100~Myr to 1~Gyr from eleven calculations, systems with 
these masses appeared to be unstable. In an analysis of three ongoing 
calculations, trends in the evolution of the $e_g$ and $a_g$ for Styx and 
Kerberos suggested these systems are also unstable. With likely 
$\tau_i \lesssim$ 2~Gyr, heavy satellite systems with the nominal masses 
derived from HST observations are unstable on time scales much smaller 
than the age of the solar system.

\citet{kb2019b} also examined the evolution of `light' satellite systems 
with
the masses listed in Table~\ref{tab: init}. 
Calculations with $f$ = 2 (1.5) 
had median $\tau_i \approx$ 100~Myr (600~Myr). Several additional complete 
simulations for the present paper result in $\tau_i \lesssim$ 3~Gyr for all 
systems with $f$ = 1.5. For models with $f$ = 1.25, two calculations had
$\tau_i$ = 700--850~Myr; trends in the evolution of $r$ with time suggested
intact systems would be unstable on time scales of 3--4~Gyr. Although all 
but one of the calculations with $f$ = 1.0 had completed 1~Gyr of evolution 
with no ejections, the evolution of the orbits of Styx and Kerberos suggested 
some were unstable.

Since the publication of \citet{kb2019b}, the completion of many additional
calculations improves the constraints on 
heavy satellite systems. All sixteen calculations with $f$ = 1 eject
at least one satellite on time scales $\lesssim$ 2~Gyr.

For light satellite systems, we divide calculations into three groups:
(i) at least one satellite has been ejected,
(ii) at least one satellite has $e \gtrsim$ 0.01 or $\imath$ $\gtrsim$ 0.01 
without an ejection, and
(iii) all satellites have $e$ and $\imath$ close to their `nominal'
values and none have been ejected. The separation into the second and 
third groups is based on a set of calculations with $f$ = 0.5 where
the satellites show no evidence of perturbations over 1~Gyr of evolution.
Within this set, the time variation of the inclination is very small:
$\imath$ = 0.00406--0.00485 with an average $\imath$ = 0.00445 for Styx and
$\imath$ = 0.00764--0.00787 with an average $\imath$ = 0.00776 for Kerberos.
Variations in eccentricity are only somewhat larger:
$\efree \lesssim$ 0.004 with an average $\efree$ = 0.002 for Styx and
$\efree$ = 0.002--0.004 with an average $\efree$ = 0.003 for Kerberos. 
Limits on the eccentricity from \eg\ and \egeo\ are similar.  Orbits of Styx
or Kerberos with $e \gtrsim$ 0.01 and $\imath \gtrsim$ 0.01 are not consistent 
with current observational limits from HST.  We therefore reject these
calculations (and their $f$ values) as possible matches to the \pc\ system.

\begin{figure}[t]
\begin{center}
\includegraphics[width=4.5in]{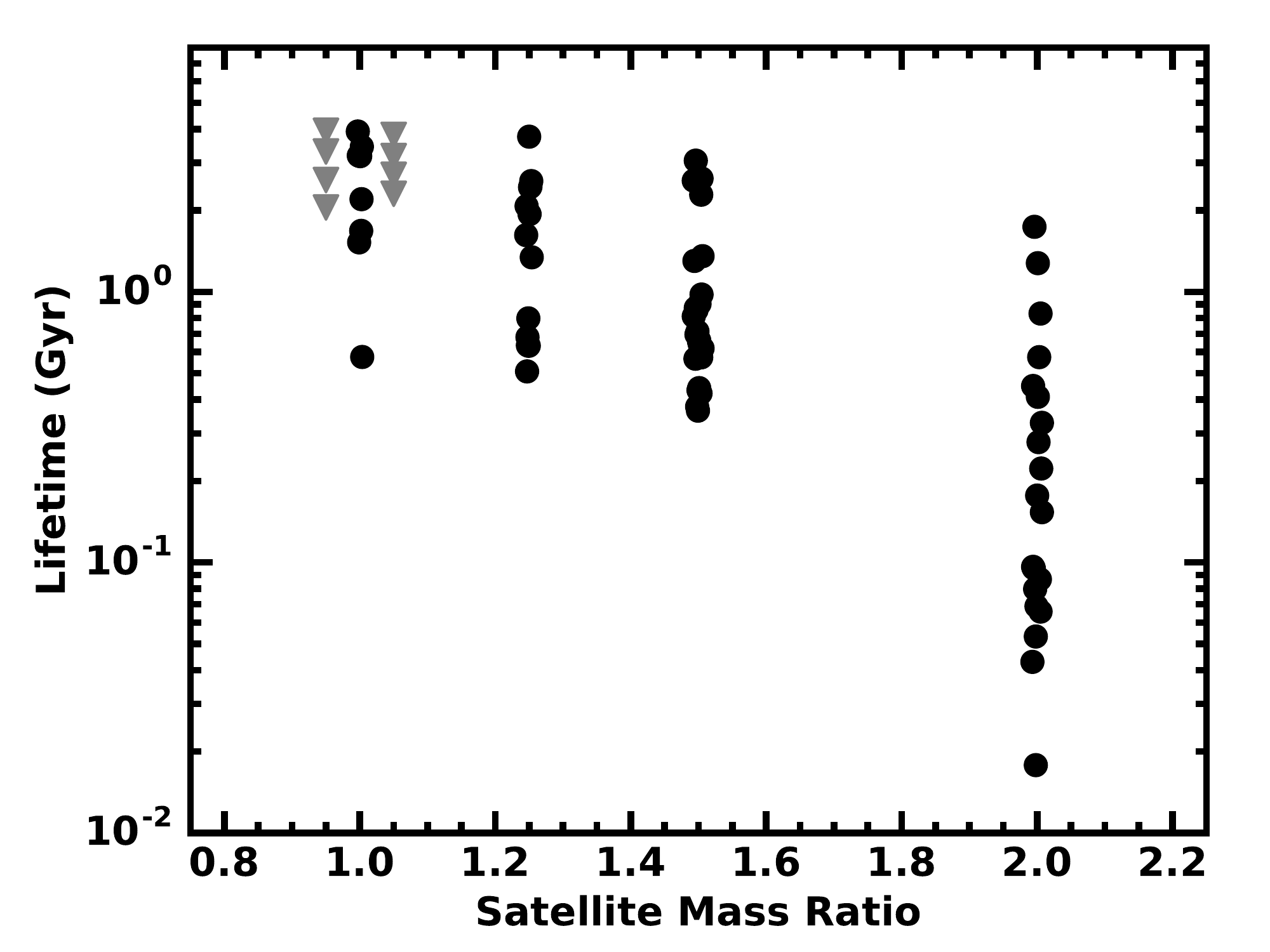}
\vskip -2ex
\caption{
\label{fig: life1}
Lifetime of model \pc\ satellite systems as a function of the mass ratio 
factor $f$. Satellites have masses $f$ times larger than the nominal 
masses listed in Table~\ref{tab: init}.
Black dots indicate lifetimes for systems with ejections of either 
Styx or Kerberos. Grey triangles represent current lifetimes for eight 
ongoing calculations with $f$ = 1.0,
which we include for completeness. 
Some points have been adjusted 
vertically or horizontally for clarity. More massive satellite systems 
typically have shorter lifetimes.
}
\end{center}
\end{figure}

With this definition, the full set of completed \nbody\ calculations yield 
strong limits on 
the combined satellite mass (Fig.~\ref{fig: life1}).  Within the suite of 18 
simulations for $f$ = 2, lifetimes range from a minimum of $\tau_i \approx$ 20~Myr 
to a maximum of $\tau_i \approx$ 2~Gyr, with a median, $\tau_i \sim$ 170~Myr, 
roughly midway between the two extremes.  
Model systems with smaller $f$ have a smaller range in $\tau_i$. Among the 
17 completed calculations with $f$ = 1.5, the maximum $\tau_i \sim$ 3~Gyr 
is only an order of magnitude larger than the minimum $\tau_i \sim$ 
400~Myr.  When $f$ = 1.25, $\tau_i$ spans 550~Myr to 3.5~Gyr among 16
calculations.  As of this writing, 8 of 16 calculations with $f$ = 1.0 have 
measured $\tau_i$ = 0.6~Myr to 3.7~Gyr.  Among the eight unfinished calculations 
that have reached $\sim$ 2--4 Gyr, none have $e_S$ or $e_K$ $\gtrsim$ 0.01.
Thus half of the calculations with $f$ = 1 are consistent with HST observations;
the other half are not.

Based on calculations with at least one ejection, it is not possible to predict 
outcomes of the ongoing calculations with $f$ = 1. Because Kerberos rarely signals
an impending ejection (Fig.~\ref{fig: idelay}), calculations with current lifetimes
of 3--4~Gyr have time to eject Kerberos before 4.5~Gyr (when we plan to end each
calculation). In roughly half of systems where Styx is ejected, Styx signals the
outcome 0.1--0.5~Gyr before an ejection. Several ongoing calculations with $f$ = 1 
have reached 3.5--4 Gyr; thus, there is time for a Styx ejection in these calculations.

\subsection{Constraints on the Masses of Nix and Hydra}

In addition to sets of calculations with masses $f$ times larger than the
nominal masses, we performed calculations where either Nix or Hydra has a
mass $f_N$ or $f_H$ times its nominal mass and other satellites have their
nominal masses. \citet{kb2019b} reported results for completed calculations
with $(f_N, f_H)$ = 1.1--5 in steps of 0.1. In this range, 77 calculations
produced an ejection; another 19 systems had evolved for at least 1~Gyr 
without an ejection. 
For each of the ongoing calculations, \citet{kb2019b}
used the time variation of \ag\ and \eg\ for Styx and Kerberos to estimate 
the likely stability of the system over 4.5~Gyr. 
Only one ongoing calculation was deemed stable.

\begin{figure}[t]
\begin{center}
\includegraphics[width=4.5in]{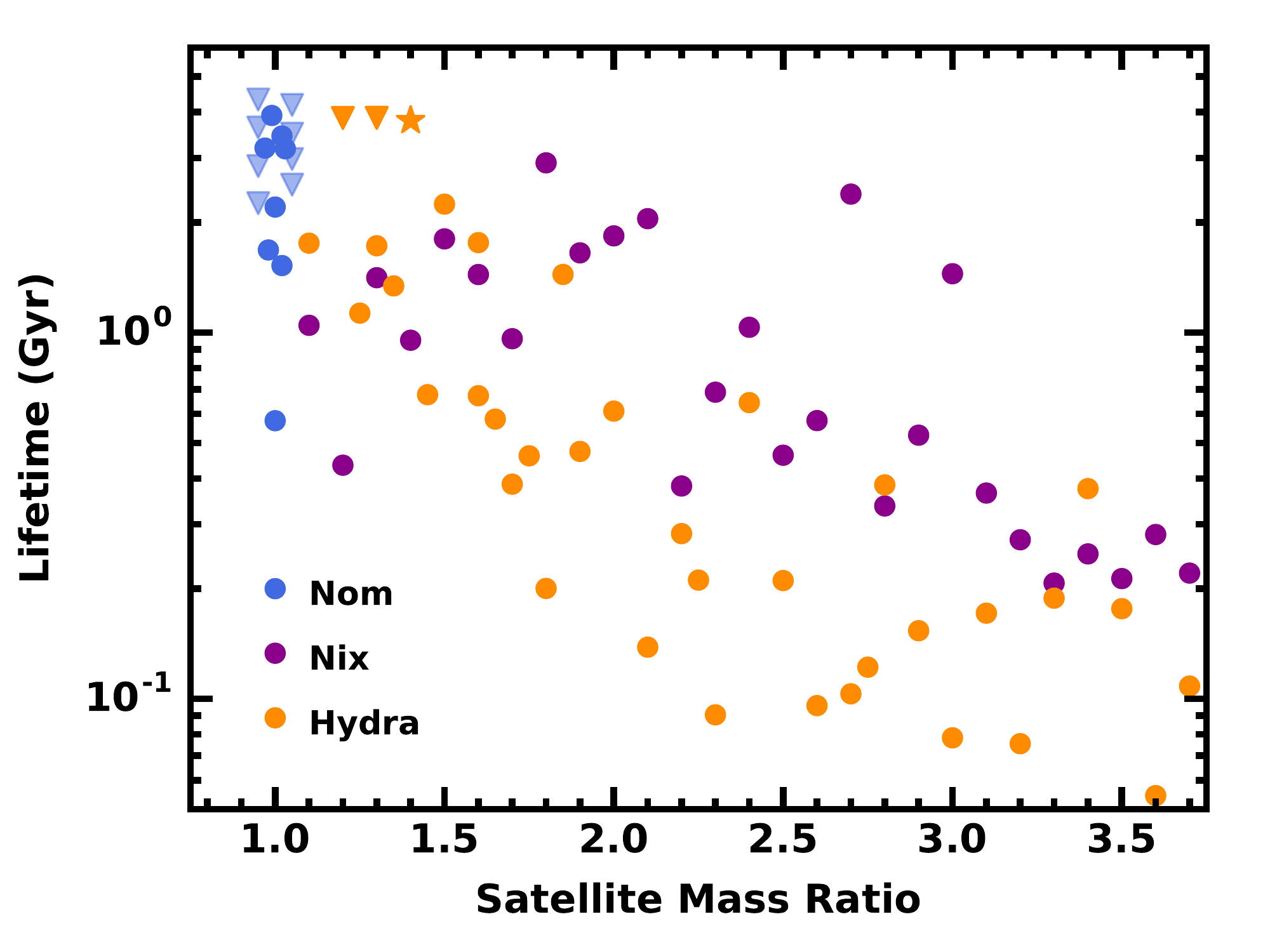}
\vskip -2ex
\caption{
\label{fig: life2}
As in Fig.~\ref{fig: life1} for model \pc\ satellite systems with
a factor $f_{N,H}$ times the nominal mass for either Nix ($f_N > 1$, 
purple dots) or Hydra ($f_H > 1$, orange dots) and nominal masses 
for the other satellites. 
Blue dots indicate lifetimes of systems with the nominal masses for all 
satellites and one or more ejections. Orange stars indicate 
systems with $f_H > 1$ where Kerberos has $e$ and $\imath$ much 
larger than derived from HST observations.  For completeness, we
include blue (orange) triangles to represent the range of lifetimes 
for eight (one) ongoing calculations with the nominal masses for all 
four satellites (with the nominal masses for Styx, Nix, and Kerberos 
and $f_H$ times the nominal mass for Hydra) where the system parameters 
still match orbital fits to HST observations.  More massive satellite 
systems have shorter lifetimes.
}
\end{center}
\end{figure}

Among the ongoing calculations in \citet{kb2019b}, all but three have
resulted in an ejection (Fig.~\ref{fig: life2}). With these results, we 
improve limits on the masses of Nix and Hydra. Of the 10 unfinished 
calculations with $f_N > 1$ and $f_H$ = 1 from \citet{kb2019b}, all 
resulted in an ejection; the maximum lifetime is $\tau_N \lesssim$ 2.7~Gyr. 
Thus, the nominal mass listed in Table~\ref{tab: init} provides a robust 
upper limit to the mass of Nix.

Of the nine previously
unfinished calculations with $f_N$ = 1 and $f_H > 1$, six resulted in an 
ejection. The range in lifetimes is 1--2.1~Gyr. In this set, the three 
unfinished calculations have $f_H$ = 1.2, 1.3, and 1.4 and evolution times 
of 3.6--3.8~Gyr. 
In two of these three ($f_H$ = 1.2 and 1.3, indicated by the filled 
orange triangles in Fig.~\ref{fig: life2}), 
Kerberos has a steadily increasing $e \gtrsim$ 0.01; $\imath$ is also larger 
than 0.01. Thus, these systems provide poor matches to the orbital elements of 
the current \pc\ satellite system. The rates of change of $e$ and $\imath$ in
these calculations suggest ejections will occur before 4.5~Gyr.

In the calculation with $f_H$ = 1.4 (Fig.~\ref{fig: life2}, filled orange star),
the orbital eccentricities and inclinations of Styx and Kerberos vary more
than calculations with $f$ = 0.5, but they do not reach the level of 0.01 that
would be much too large to match the observed orbital elements. Thus, this system
might end up matching the real satellite system. Even if this model system remains
stable, the mass of Hydra is unlikely to exceed its nominal mass. With eight
ejections for $f$ = 1, ejections for $f_H$ = 1.1 and $\gtrsim$ 1.5, and poor 
matches to the real \pc\ system for $f_H$ = 1.2 and 1.3, the most likely mass
for Hydra is less than or equal to its nominal mass.

\subsection{Constraints on the Masses of Styx and Kerberos}

To try to place initial constraints on the masses of Styx and Kerberos,
we vary their masses independently of the masses of Nix and Hydra. For
each calculation, we adopt $f_S$ = $f_K$ = 1.0, 1.5, 2.0, or 3.0 and
then adopt $f$ = 1.0, 1.25, 1.5, 2.0, 2.5, and 3.0 for the full set of 
four satellites. 
As an example, when $f$ = 2 and $f_S = f_K$ = 1.5, Nix and Hydra have
twice their nominal masses; Styx and Kerberos have thrice their nominal
masses.
Here, we discuss results for $f$ = 1.5--4.0 and defer consideration of 
calculations with $f$ = 1.0--1.25 to a later study.

The goal of this set of calculations is to learn whether the lifetimes of
satellite systems with $f_S$ = 2--3 are significantly longer than the 
lifetimes of systems with $f_S$ = 1.0--1.5. When $f$ = 2.5 or 3.0, rapid
ejections of Styx or Kerberos should make it difficult to measure different
lifetimes in systems with different masses for Styx and Kerberos. These 
calculations serve to mimimize small number statistics in deriving a median
lifetime. For smaller $f$, longer lifetimes enable tests to looks for
differences among calculations with different $f_S$ and $f_K$.

\begin{figure}[t]
\begin{center}
\includegraphics[width=4.5in]{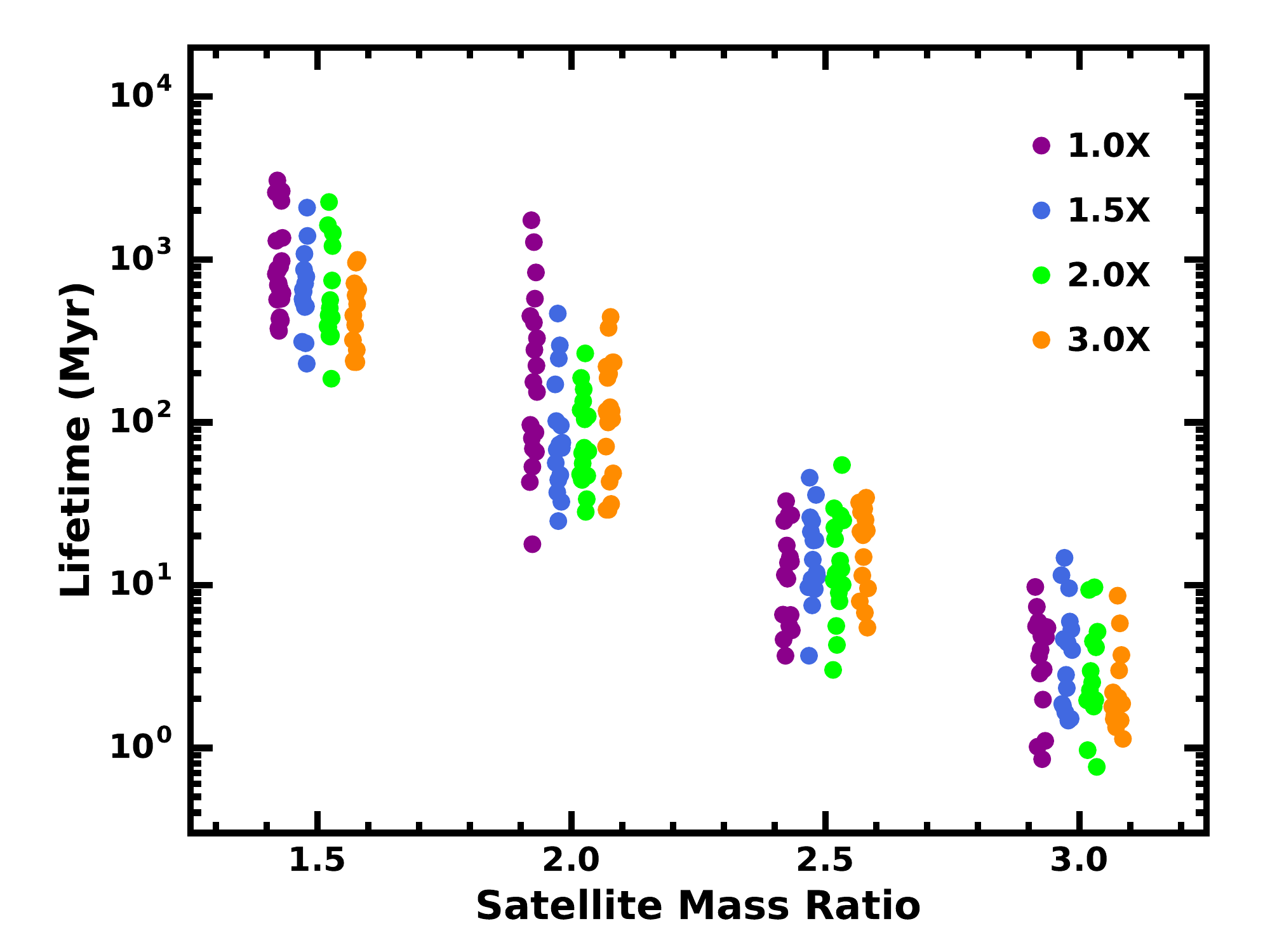}
\vskip -2ex
\caption{
\label{fig: life3}
Lifetime of model \pc\ satellite systems as a function of the 
mass ratio $f$ for different $f_{S, K}$. The legend indicates 
the adopted mass ratio $f_S = f_K$ for the masses of Styx and Kerberos 
relative to their nominal masses. For each calculation, Nix and Hydra 
have their nominal masses multiplied by the factor $f$; Styx and Kerberos
have their nominal masses multiplied by $f * f_{S, K}$. For $f$ = 1.5--2, 
lifetimes for systems with larger mass ratios for Styx and Kerberos 
(e.g., $f_{S, K}$ = 2 or 3) have shorter lifetimes.
}
\end{center}
\end{figure}

Fig.~\ref{fig: life3} plots the lifetimes of the satellite systems for
calculations with $f$ = 1.5--3.0. Within each set, the typical range 
in lifetimes is a factor of $\sim$ 10. Curiously, the range is largest for
simulations with $f$ = 2 for the nominal satellite masses. A larger set of
calculations is required to learn whether this result is due to the small
number of simulations or a real effect. For a complete suite of calculations
with fixed $f_S$ and variable $f$, the lifetimes grow with decreasing $f$. 
Visually, the lifetimes for systems with $f$ = 1.5 appear smaller for
larger $f_S$ than for smaller $f_S$.

Table~\ref{tab: SKdata} summarizes lifetimes for this suite of 
calculations. Aside from $f_S = f_K$ and $f$, the columns list 
the number of 
results for each combination of $f_S$ and $f$ and the minimum, median, and 
maximum lifetimes. Minimum lifetimes range from roughly 0.3~Myr for $f$ = 4
to 200--300~Myr for $f$ = 1.5. Maximum lifetimes are $\sim$ 10 times larger,
ranging from 2--3~Myr for $f$ = 4 to 1--3~Gyr for $f$ = 1.5. Median 
lifetimes are roughly midway between the minimum and maximum values. 

For large $f$ = 2--4, the minimum, median, and maximum lifetimes for a
specific $f_S$ are rather uncorrelated. Systems with $f_S$ = 1 and 
$f_S$ = 3 have nearly identical values. Thus, the masses of Styx and Kerberos
have little influence on the lifetimes of massive satellite systems.
When $f$ is smaller, however, there is a systematic increase in the median
and maximum lifetimes from $f_S$ = 3 to $f_S$ = 1. In these calculations, 
the masses of Styx and Kerberos clearly change the median lifetime.

To place the influence of Styx and Kerberos on a more quantitative footing,
we compare the distributions of lifetimes for combinations of $f$ and $f_s$.
We use the Python version of the Mann-Whitney–Wilcoxon rank-sum test, 
which ranks the lifetimes and uses a non-parametric technique to measure
the probability that the two distributions are drawn from the same parent
distribution. For $f$ = 2--4, the rank-sum probabilities are generally
large, $\gtrsim$ 0.3, and indicate a correspondingly large probability that
the distributions are drawn from the same parent population as suggested
from the minimum, median, and maximum lifetimes in Table~\ref{tab: SKdata}.
For calculations with $f$ = 1.5, however, the rank-sum test returns a small
probability, $p$ = 0.05, that calculations with $f_S$ = 1 and $f_S$ = 2 are 
drawn from the same population. The test returns a smaller probability,
$p$ = 0.005, that the set of lifetimes for $f$ = 1 and $f$ = 3 are drawn from
the same parent. Lifetimes for $f$ = 1 and $f$ = 1.5 have a high
probability, $p$ = 0.29, of belonging to the same parent.

As a final check on these results, we consider the Python version of
the non-parametric K--S test. This test leads to the same conclusions:
high probabilities that lifetimes with $f \ge$ 2 for all $f_S$ are drawn 
from the same parent population and low probabilities that the set of
lifetimes with $f$ = 1.5 and $f_S$ = 1 and either $f_S$ = 2 ($p$ = 0.07) or 
$f_S$ = 3 ($p$ = 0.08) are drawn from the same population.

\begin{deluxetable}{cccccc}
\tablecolumns{6}
\tablewidth{15cm}
\tabletypesize{\normalsize}
\tablenum{3}
\tablecaption{Lifetimes for Calculations with $f_S = f_K$ = 1--3\tablenotemark{1}}
\tablehead{
\colhead{$f_S, f_K$} &
\colhead{~~~$f$~~~} &
\colhead{~~~~N~~~~~} &
\colhead{~~~~~$\tau_{min}$~~~~~} &
\colhead{~~~~~$\tau_m$~~~~~} &
\colhead{~~~~~$\tau_{max}$~~~~~}
}
\label{tab: SKdata}
\startdata
1.0 & 4.0 & 15 & 13.001 & 13.251 & 13.674 \\
1.0 & 3.5 & 15 & 13.160 & 13.467 & 13.890 \\
1.0 & 3.0 & 17 & 13.430 & 14.179 & 14.487 \\
1.0 & 2.5 & 16 & 14.065 & 14.600 & 15.015 \\
1.0 & 2.0 & 19 & 14.750 & 15.685 & 16.740 \\
1.0 & 1.5 & 19 & 16.060 & 16.342 & 16.984 \\
1.5 & 4.0 & 15 & 13.047 & 13.525 & 13.839 \\
1.5 & 3.5 & 15 & 13.085 & 13.601 & 14.096 \\
1.5 & 3.0 & 15 & 13.667 & 14.100 & 14.666 \\
1.5 & 2.5 & 16 & 14.065 & 14.563 & 15.158 \\
1.5 & 2.0 & 16 & 14.892 & 15.352 & 16.167 \\
1.5 & 1.5 & 14 & 15.859 & 16.300 & 16.817 \\
2.0 & 4.0 & 15 & 12.992 & 13.370 & 13.908 \\
2.0 & 3.5 & 16 & 13.262 & 13.739 & 14.463 \\
2.0 & 3.0 & 15 & 13.382 & 13.853 & 14.485 \\
2.0 & 2.5 & 17 & 13.978 & 14.585 & 15.236 \\
2.0 & 2.0 & 18 & 14.949 & 15.315 & 15.922 \\
2.0 & 1.5 & 15 & 15.766 & 16.155 & 16.852 \\
3.0 & 4.0 & 15 & 13.068 & 13.389 & 13.637 \\
3.0 & 3.5 & 15 & 13.268 & 13.578 & 13.931 \\
3.0 & 3.0 & 15 & 13.554 & 13.771 & 14.434 \\
3.0 & 2.5 & 15 & 14.237 & 14.828 & 15.036 \\
3.0 & 2.0 & 18 & 14.961 & 15.568 & 16.146 \\
3.0 & 1.5 & 14 & 15.868 & 16.127 & 16.496 \\
\enddata
\tablenotetext{1}{The columns list $f_{S, K}$, $f$, the number of
completed calculations $N$, and the minimum ($\tau_{min}$),
median ($\tau_m$), and maximum ($\tau_{max}$) lifetimes among the
$N$ calculations for each combination of $f$ and $f_{S, K}$.
}
\end{deluxetable}

Taken together, the results listed in Table~\ref{tab: SKdata} and from
the K--S and rank-sum tests indicate that satellites systems with $f_S$ =
2--3 are more unstable than those with $f_S$ = 1.0--1.5. Because systems
with the nominal masses and $f$ = 1 are barely stable, we conclude that
lifetimes derived from the \nbody\ calculations are sufficient to place
limits on the masses of Styx and Kerberos, despite their small masses 
compared to Nix and Hydra. If this trend continues with the calculations
for $f$ = 1.25 and $f$ = 1.0, then it should be possible to rule out systems
where Nix and Hydra have their nominal masses but Styx and Kerberos are 2--3
times more massive than their nominal masses.

For this study, we conclude that 
Styx and Kerberos are more likely to have masses $\lesssim$ 1.5 times their 
nominal masses than $\gtrsim$ 2 times their nominal masses. For the dimensions 
measured from \nh\ \citep{weaver2016}, the average bulk densities for Styx 
and Kerberos 
are then $\rho_{SK} \lesssim$ 1.5~\gcmc, similar to the derived 
average bulk densities of Nix and Hydra \citep{kb2019b}.

Although stable satellite systems with smaller masses for Nix and Hydra 
{\it and} larger masses for Styx and Kerberos are possible, this option 
seems unlikely. Reducing the masses of Nix and Hydra below the nominal 
masses yields average bulk densities $\rho_{NH} \lesssim$ 1.25--1.30~\gcmc.
Allowing $f_S \approx f_K \gtrsim$ 2 establishes much larger 
average bulk densities
for the smallest satellites, $\rho_{SK} \gtrsim$ 2~\gcmc. If all four 
satellites grow out of debris from a giant impact \citep[e.g.,][]{canup2005,
canup2011,arakawa2019,bk2020,kb2021a}, it is unclear why they should have such 
different bulk densities. Thus, we conclude that the combined system mass
$m_{SNKH} \lesssim 9.5 \times 10^{19}$~g favors bulk densities for Styx
and Kerberos similar to those of Nix and Hydra.

\section{Discussion} \label{sec: disc}

Together with 
measurement of $(e_S, \imath_S)$ and $(e_K, \imath_K)$ of several ongoing
calculations,
the $\sim$ 1200 completed \nbody\ calculations described here and in
\citet{kb2019b} paint a clear picture for the masses of the four small 
satellites of \pc. The marginal stability of systems with the nominal 
masses and the instability in nearly all
systems with the nominal
masses and either $f_N >$ 1 or $f_H >$ 1 set a strict upper limit on the
combined mass of all four satellites, 
$M_{SNKH} \lesssim 9.5 \times 10^{19}$~g. The rank-sum and KS tests 
indicate the masses of Styx and Kerberos are no more than twice their 
nominal masses. If these two small satellites have masses larger than 
their nominal masses, the masses of Nix and Hydra must be reduced to 
maintain the upper limit on the total system mass.

In \citet{kb2019b}, 
we derived probability distributions for the satellite bulk density 
with a Monte Carlo (MC) calculation.
From \nh, the satellites have measured sizes and associated errors in
three dimensions, e.g., $x_k \pm \delta x_k$, for three principal axes 
$k$ = 1, 2, 3 \citep{weaver2016}. 
Defining the volume as some function of the dimensions, 
$V = f(x_k)$, and assuming a gaussian distribution for the measurement
errors, the MC analysis yields a probability distribution for the volume,
e.g., $p(V_n)$, where $n$ = 1--N is a Monte Carlo trial and $N = 10^4$. 
For an adopted satellite mass $m_n$, the bulk density for each trial is 
$\rho_n = m_n / V_n$. We report the median of the bulk
density distribution and adopt the inter-quartile range as a measure of
the uncertainty in the median. \citet{kb2019b} also described results
for the bulk density derived from a probability distribution of satellite
masses, $p(m_n)$. Each MC trial then samples $p(m_n)$ and $p(V_n)$ to
infer a probability distribution for $\rho_n$. 

Compared to \citet{kb2019b}, we have stronger limits on the satellite
masses for $f \ge 1$ and no additional information on masses for $f < 1$.
Thus, we derive bulk density estimates for the nominal masses in 
Table~\ref{tab: init}. Although \citet{kb2019b} considered two options 
for errors in the size measurements from \nh, the bulk densities were
fairly independent of plausible choices. We refer readers to Table~3 of
\citet{kb2019b} for bulk density estimates derived from adopted mass 
distributions and different choices for size errors.

\citet{kb2019b} considered three possible shapes for the small satellites,
boxes, triaxial ellipsoids, and pyramids. However, satellite images from 
\nh\ do not resemble boxes (where the volume is close to a maximum) or 
pyramids (where the volume is close to a minimum). Triaxial ellipsoids, 
where the volume $V = 4 \pi x_1 x_2 x_3 / 3$, are plausible shapes and 
have a volume intermediate between boxes and pyramids. For \nh\ images of
the Kuiper Belt object (486958) Arrokoth, the best-fitting global shape 
model of each lobe closely resembles triaxial ellipsoids 
\citep{spencer2020}. Thus, our choice for the \pc\ satellites is 
reasonable. 

With no new analysis of \nh\ images, we rely on the 
\citet{kb2019b} MC estimate for the volumes of Nix and Hydra. Scaling
the results for the improved limits on the mass yields a median bulk
density, $\rho_N \approx$ 1.55~\gcmc\ for Nix and 
$\rho_H \approx$ 1.30~\gcmc\ for Hydra. Using the interquartile range,
the error in the bulk density is $\pm$0.2~\gcmc\ for Nix and 
$\pm$0.4~\gcmc\ for Hydra. The large error for Hydra is a result of larger
errors in the measured dimensions from \nh. The probability that the 
satellites have smaller bulk densities than Charon (Pluto) is 
65\% (80\%) for Nix and 75\% (90\%) for Hydra. Hydra's lower bulk density 
results in a higher probability of a bulk density that is lower than 
Charon's or Pluto's.

Given the uncertainties, the average bulk density of the four 
satellites is comparable to the average bulk density of Nix and Hydra, 
$\rho_{SNKH} \lesssim$ 1.4~\gcmc. Plausible errors in this estimate are
$\pm$0.1--0.2~\gcmc. The shorter lifetimes of satellite systems with
$f_{S, K} \gtrsim$ 2 point to similar bulk densities 
$\rho_{SK} \lesssim$ 1.5~\gcmc\ for Styx and Kerberos. 

The bulk densities derived for the \pc\ satellites are similar to the
densities of other small objects in the solar system. Satellites of 
Mars, Saturn, and Uranus with similar sizes as Styx/Nix/Kerberos/Hydra
have $\rho \approx$ 0.5--1.5~\gcmc\
\citep[e.g.,][]{jacobson2004,renner2005,jacobson2010,patz2014,chancia2017}.
Curiously, Kuiper belt objects (KBOs) with much larger
radii, $r \approx$ 50--200~km, have much lower bulk densities, $\sim$ 
0.5--1.0~\gcmc; larger
KBOs with $r \approx$ 500--1000~km have higher bulk densities, $\gtrsim$
1.5~\gcmc\ \citep[e.g.,][]{brown2012,grundy2019}. 
Bulk density measurements of other satellites are needed to place the 
\pc\ satellites in context with moons throughout the solar system.

In the inner solar system, high quality astrometric data provide evidence 
for rubble-pile structures in some asteroids \citep[e.g.,][]{chesley2014}.
Coupled with high quality modeling, these data suggest low bulk densities
and high porosity. \citet{bierson2019} demonstrate that the variation in
bulk density of KBOs could result from a variation in porosity, where
smaller KBOs have a much larger porosity than larger KBOs. In this 
scenario, the additional mass of larger KBOs compacts the structure and
generates a smaller porosity.

In the \pc\ satellites, the bulk density estimates are based on an upper
limit of the mass and a median volume derived from a probability
distribution, $\rho = m / V_{med}$. In this approach, allowing for 
empty regions in each satellite has no impact on the overall bulk density.
The derived masses and volumes are the same. However, a high degree of 
void in a satellite increases the bulk density of the non-void material. 
As an example, the porosity required for Pluto-like material with 
$\rho_P$ = 1.85~\gcmc\ to have the bulk density of Nix, 
$\rho_N \approx$ 1.55~\gcmc, is 16\%. For Hydra, the required porosity 
is 30\%. In both satellites, the porosity required for non-void material
to have the bulk density of Charon is smaller, $\sim$ 9\% for Nix and
$\sim$ 24\% for Hydra. Porosities much larger than these estimates require
satellite compositions dominated by rock, which seems unlikely.

Even with uncertainties regarding the porosity of the small satellites,
the results described here provide stronger support that the satellites 
are icy material ejected 
(i) from Pluto during a giant impact that results in a binary
planet \citep[e.g.,][]{canup2005,canup2011,kb2021a,canup2021} or 
(ii) from a more modest impact
between a trans-Neptunian object and Charon \citep{bk2020}. In models where
Charon forms out of the debris from the giant impact 
\citep[e.g.,][]{desch2015,desch2017}, the bulk densities of the small 
satellites are more likely to be similar to Charon than to the low densities
derived here. In \nbody\ simulations of Charon formation, dynamical 
interactions are unlikely to leave behind sufficient material for the small
satellites \citep{kb2019c}.

In addition to limits on satellite masses and bulk densities, 
the suite of \nbody\ 
calculations provide interesting information on the frequency of 
ejections and the prelude to an ejection. In systems with at least twice
the nominal mass, Kerberos is ejected much more often than Styx. For
lower mass systems, Styx ejections are more likely. Within the suite of 
calculations for systems with $f \le$ 2, Styx often signals an upcoming
ejection by developing a relatively large inclination, $\imath_S \gtrsim$
0.01. Styx can remain at this inclination many Myr before an ejection.

This behavior is a natural outcome of the orbital architecture of 
the four small satellites. Because Kerberos--Styx has the smallest 
$K = (a_i - a_j) / R_{H, ij}$, it is the most likely satellite to suffer
orbital perturbations from Nix and Hydra. Kerberos is also the closer
of the two smaller satellites to an orbital resonance with the central
binary. \citet{youdin2012} demonstrated that the 5:1 resonance with
\pc\ is unstable. Thus it is not surprising that Kerberos is more likely
to be ejected than Styx in massive satellite systems.

The behavior of Styx in the \nbody\ calculations is fascinating. Despite
having a larger $K$ and orbiting farther away from an orbital resonance,
it is often ejected in low mass satellite systems. Its ability to signal 
an ejection while maintaining a modest $\imath \approx$ 0.02--0.03 for
several hundred Myr is a consequence of angular momentum conservation:
when perturbations from Nix and Hydra increase $e_S$, the satellite is
able to reduce $e$ (thus reducing perturbations) by raising $\imath$ and
maintaining stability. Because Kerberos has a more precarious orbit, it
does not have this option and rarely signals an upcoming ejection.

\section{Summary} \label{sec: summary}
 
We analyze a new suite of $\sim$ 500 numerical \nbody\ calculations to 
constrain the masses and bulk densities of the four small satellites of 
the \pc\ system. To infer the semimajor axis and eccentricity of 
circumbinary satellites from the six phase-space coordinates from the
\nbody\ code, we consider four approaches. 
Keplerian elements derived from the orbital energy and angular momentum 
(eqs.~\ref{eq: akep}--\ref{eq: ekep}) poorly represent circumbinary orbits. 
Two geometric estimates (eqs.~\ref{eq: ag}--\ref{eq: egeo}) enable good 
results for long \nbody\ integrations but also require good sampling over 
many orbits \citep[see also][]{sutherland2019}. Geometric estimates based 
on restricted three-body theory yield somewhat smaller and more 
accurate measures of $a$ and $e$ for the \pc\ satellites. An instantaneous
estimate derived from the restricted three-body problem provides the
radius of the guiding center \rgc\ as a surrogate for $a$ and the free
eccentricity \efree. For the \pc\ satellites (especially Styx and Nix), 
\rgc\ is a better measure of the semimajor axis than \ag\ or \ageo; 
\efree\ and \egeo\ are roughly equivalent measures of the eccentricity.

Results from the new calculations build on the analysis of $\sim$ 700
simulations from \citet{kb2019b}. The earlier calculations demonstrated
that heavy satellite systems -- where the masses of Styx and Kerberos 
are much larger than those in Table~\ref{tab: init} \citep{brozovic2015} --
are unstable. Another set of early calculations showed that light satellite
systems with masses $f \ge $ 1.5 times larger than the nominal masses in 
Table~\ref{tab: init} are also unstable. Finally, a third set of results
yielded robust upper limits on the masses of Nix ($\le$ twice the nominal 
mass) and Hydra ($\le$ 1.5 times the nominal mass).

The analysis described here focuses solely on light satellite systems. 
Completed calculations with mass fractions (i) $f$ = 1.00--1.25 times 
the nominal mass of the combined satellite system, (ii) $f_N$ = 1--2
times the mass of Nix, and (iii) $f_H$ = 1.0--1.5 times the mass of Hydra 
place better constraints on the total system mass. 
We also derived new results for systems with 1.5--4 times the nominal 
masses of Nix and Hydra and 2.25--12 times the nominal masses of Styx 
and Kerberos to understand whether system lifetimes depend on the masses
of the two smallest satellites.

When combined with the $\sim$ 700 simulations from \citet{kb2019b}, we 
draw the following conclusions.

\begin{itemize}

\item 
When the mass of the satellite system is more than twice the nominal
mass (Table~\ref{tab: init}, Kerberos is ejected more often than Styx. 
In lower mass systems, Styx is more likely to be lost than Kerberos. 
In either case, ejections occur when Nix or Hydra (or both) perturb the 
orbit of Styx or Kerberos across an orbital resonance with the central 
binary. The central binary, Nix, and Hydra then drive the satellite 
beyond the innermost stable orbit.  The central binary then ejects the 
wayward satellite from the \pc\ system.

\item 
When the inclination of one of the smaller satellites rises above 0.01,
it `signals' an impending ejection. The signals of Kerberos are rather
weak and often occur only a few Myr before ejection. Styx often signals
strongly several tens or hundreds of Myr before an ejection.

\item Satellite systems with the nominal masses listed in 
Table~\ref{tab: init} are marginally stable. The set of completed
calculations yields a robust upper limit on the mass of the combined 
satellite system, $m_{SNKH} \lesssim 9.5 \times 10^{19}$~g. 
Although this mass estimate is nearly identical to the estimate in 
\citet{kb2019b}, the present result is based on a larger set of 
completed calculations with satellite masses close to the nominal 
masses in Table~\ref{tab: init}.
Adopting a triaxial ellipsoid model for the shape of each satellite, 
the satellite dimensions measured by \nh\ and the upper limit on the 
combined mass implies an average bulk density of $\rho_{SNKH}$ =
1.25~\gcmc, which is significantly smaller than the bulk density of 
Charon and Pluto.

\item Calculations where the masses of Styx and Kerberos are 2--3 times
larger than their nominal masses have significantly shorter lifetimes than
calculations where the masses are $\lesssim$ 1.5 times the nominal masses. 
This result indicates that the bulk densities of Styx and Kerberos are 
probably
closer to the bulk density of ice than to the bulk density of rock. An 
icy composition is consistent with the large measured albedos of both 
satellites.

\end{itemize}

Improved constraints on the bulk densities of the four small satellites
require better limits on the masses and the volumes 
\citep[see also][]{canup2021}.  
Completion of \nbody\ calculations with
$f$ = 0.5--0.875 will establish a robust set of the masses required for
stable satellite systems.
Another set with $f$ = 1.00--1.25 for Nix/Hydra and $f$ = 1.5--4.75 for
Styx/Kerberos will yield better estimates of the masses for Styx and Kerberos.
Choosing among possible volume estimates requires more detailed shape models, 
as in studies of Arrokoth \citep[e.g.,][]{mckinnon2020a,
spencer2020,stern2021}. Together, these modeling efforts will enable a
clearer picture of the origin and early evolution of the \pc\ satellite
system.

\vskip 6ex

We acknowledge generous allotments of computer time on the NASA `discover' 
cluster, provided by the NASA High-End Computing (HEC) Program through 
the NASA Center for Climate Simulation (NCCS).
Advice and comments from M. Geller and two anonymous reviewers
improved the presentation.
Portions of this project were supported by the {\it NASA } {\it Outer 
Planets} and {\it Emerging Worlds} programs through grants NNX11AM37G and 
NNX17AE24G.
Some of the data (binary output files and C programs capable of reading 
them) generated from this numerical study of the \pc\ system are available 
at a publicly-accessible repository (https://hive.utah.edu/) with url 
https://doi.org/10.7278/S50d-5g6f-yfc5.

\bibliography{ms.bbl}{}
\bibliographystyle{aasjournal}

\end{document}